\documentclass[prl,aps,twocolumn,superscriptaddress,fleqn]{revtex4-2}

 \usepackage{
 	graphicx,
 	subfigure,
 	placeins
 }
 \graphicspath{{Figures/}}
 
 \usepackage{graphicx,subfigure,placeins,wrapfig}
 
 \usepackage[english]{babel}
 \usepackage{blindtext}
 
 \usepackage[utf8]{inputenc}
 \usepackage[T1]{fontenc}
 \usepackage{xcolor,booktabs,color,lipsum,makecell,microtype,fix-cm}
 \usepackage[normalem]{ulem}
 \usepackage{bm, amsmath,float,amssymb,upgreek,physics,braket,mhchem}
 \usepackage[exponent-product=\cdot]{siunitx}

\newcommand{\bl}[1]{\textcolor{black}{#1}}

\usepackage{nicefrac} 			

\usepackage{tikz}

\usepackage{}

\newcommand{\Vbg}{V_\mathrm{bg}}
\newcommand{\nUx}{U_x/P_\mathrm{s}}
\newcommand{\nUy}{U_y/P_\mathrm{s}}
\newcommand{\Vp}{V_\mathrm{p}}
\usepackage{chngcntr,printlen}
\usepackage[unicode=true,colorlinks=true,citecolor=blue,urlcolor=blue]{hyperref}

\begin{document}
 	
 	\title{Terahertz and gigahertz magneto-ratchets in graphene-based 2D metamaterials}
 	
 	\author{M. Hild}
 	\affiliation{Physics Department, University of Regensburg, 93040 Regensburg, Germany}
 	
 	\author{E. Mönch}
 	\affiliation{Physics Department, University of Regensburg, 93040 Regensburg, Germany}
 	
 	\author{L. E. Golub}
 	\affiliation{Physics Department, University of Regensburg, 93040 Regensburg, Germany}
 	
 	\author{I. A. Dmitriev}
 	\affiliation{Physics Department, University of Regensburg, 93040 Regensburg, Germany}
 	
 	\author{I. Yahniuk}
 	\affiliation{Physics Department, University of Regensburg, 93040 Regensburg, Germany}
 	
 	\author{K. Amann}
 	\affiliation{Physics Department, University of Regensburg, 93040 Regensburg, Germany}
 	
 	\author{J. Amann}
 	\affiliation{Physics Department, University of Regensburg, 93040 Regensburg, Germany}
 	
 	\author{J. Eroms}
 	\affiliation{Physics Department, University of Regensburg, 93040 Regensburg, Germany}
 	
 	\author{J. Wunderlich}
 	\affiliation{Physics Department, University of Regensburg, 93040 Regensburg, Germany}
 	
 	\author{D. Weiss}
 	\affiliation{Physics Department, University of Regensburg, 93040 Regensburg, Germany}
 	
 	\author{C. Consejo}
 	\affiliation{L2C UMR 5221, Université de Montpellier, CNRS, 34090 Montpellier, France}
 	
 	\author{C. Bray}
 	\affiliation{L2C UMR 5221, Université de Montpellier, CNRS, 34090 Montpellier, France}
 	
 		\author{K. Maussang}
 	\affiliation{Institut d’Electronique et des Systèmes, UMR5214, Université de Montpellier, CNRS, 34000 Montpellier, France}

 	\author{F. Teppe}
 	\affiliation{L2C UMR 5221, Université de Montpellier, CNRS, 34090 Montpellier, France}
 	
 	\author{J. Gumenjuk-Sichevska}
 	\affiliation{Johannes Gutenberg-University Mainz, D-55128 Mainz, Germany }
 	\affiliation{V. Lashkaryov Institute of Semiconductor Physics, National Academy of Science, 03028, Kyiv, Ukraine }

 	\author{K. Watanabe}
 	\affiliation{Research Center for Electronic and Optical Materials, National Institute for Materials Science, 1-1 Namiki, Tsukuba 305-0044, Japan}
 	
 	\author{T. Taniguchi}
 	\affiliation{Research Center for Materials Nanoarchitectonics,  National Institute for Materials Science, 1-1 Namiki, Tsukuba 305-0044, Japan}
 	
 	\author{S. D. Ganichev}
 	\affiliation{Physics Department, University of Regensburg, 93040 Regensburg, Germany}
 	\affiliation{CENTERA Labs, Institute of High Pressure Physics, PAS, 01 - 142 Warsaw, Poland}

 	\begin{abstract}
 		We report on the observation and study of the magneto-ratchet effect in a graphene-based two-dimensional metamaterial formed by a graphite gate that is placed below a graphene monolayer and patterned with an array of triangular antidots. We demonstrate that terahertz/gigahertz excitation of the metamaterial leads to sign-alternating magneto-oscillations with an amplitude that exceeds the ratchet current at zero magnetic field by orders of magnitude. The oscillations are shown to be related to the Shubnikov-de Haas effect. In addition to the giant ratchet current oscillations we detect resonant ratchet currents caused by the cyclotron and electron spin resonances. The results are well described by the developed theory considering the magneto-ratchet effect caused by the interplay of the near-field radiation and the nonuniform periodic electrostatic potential of the metamaterial controlled by the gate voltages.
 	\end{abstract}
 	
 	\maketitle

 	\section{Introduction}

 	\label{introduction}

 	Artificial periodic metallic/dielectric structures with a period smaller than the radiation wavelength fabricated on top/bottom of graphene have attracted increasing research interest in the last decade. This is due to their extraordinary ability to manipulate electromagnetic (EM) waves and excite various physical phenomena that are not available in unstructured graphene layers. Metamaterials designed for terahertz (THz) frequencies are of particular interest due to their high application potential. Some examples are perfect absorbers based on near-field effects, plasmonic devices, nonlinear optics, high-speed modulators, biosensing etc., see Refs.\,\cite{Caldwell2015, Xu2017b, Nemati2018, Yu2018b, He2019, Xiao2020, Xu2022}. Direct currents excited by THz electric fields in graphene-based asymmetric structures formed by periodically repeating metal stripes of different widths deposited on mono- and bilayer graphene layers have been proposed as promising detectors of THz radiation. They have been intensively studied over a wide range of frequencies from hundreds of gigahertz (GHz) to tens of THz~\cite{Olbrich2016,Ganichev2017,Fateev2017,Fateev2019,BoubangaTombet2020,DelgadoNotario2020,DelgadoNotario2022,Morozov2021,Moench2022,Moench2023}. Important established mechanisms for converting THz/GHz radiation into direct currents (dc) are electronic and plasmonic ratchet effects.

 		\begin{figure*}[t] 
 		\centering
 		\includegraphics[width=\linewidth]{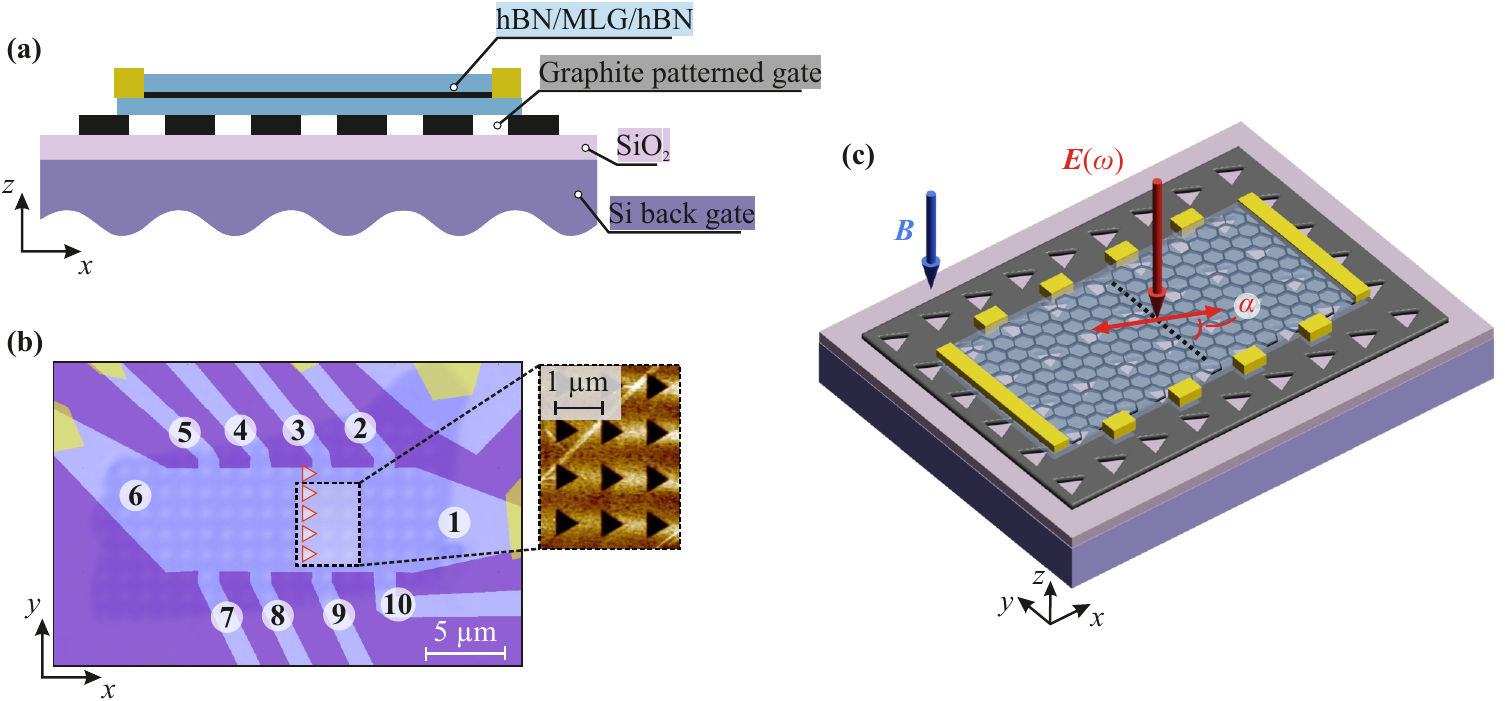}
 		\caption{Sketch of the sample and the experimental setup. Panel (a): The cross section of the sample. Panel (b): Photograph of sample~\#A (left) and an AFM image of the periodically patterned gate (right). The patterned gate made of five graphene layers has equilateral triangular holes arranged in a square lattice. In the photo, a row of triangles is highlighted by red color. The $x$-direction is along the height of the triangles, and the $y$-direction is along their base. Panel (c): Experimental configuration. The red vertical arrow shows the normal incident linearly polarized radiation with a wavelength much larger than the lattice period $d$ and a laser spot much larger than the Hall bar. The red double arrow sketches the radiation electric field vector $\bm{E}(\omega)$, whose orientation with respect to the triangle height is defined by the azimuth angle $\alpha$. The blue arrow indicates the magnetic field $\bm{B}$ applied perpendicularly to the sample plane.}
 		\label{fig1}
 	\end{figure*}
 	
 	Here, we report the observation of the magneto-ratchet effect in a two-dimensional 2D graphene-based metamaterial formed by a graphite gate patterned with an array of triangular antidots placed under a graphene monolayer. \bl{Most recently we demonstrated that excitation of these metamaterials by terahertz radiation results in a 2D ratchet effect, which efficiently converts high frequency electric fields into dc electric current~\cite{Hild2024,Yahniuk2024}. Here we show that the application of an external magnetic field leads to resonant and non-resonant magneto-ratchet phenomena, including cyclotron and spin resonances, that are characterized by a giant magnitude and rich physics.} The photocurrent is excited by linearly polarized low-power continuous wave (cw) THz/GHz radiation in the presence of a magnetic field oriented perpendicular to the graphene plane. \bl{By varying the magnetic field we detected sign-alternating $1/B$-periodic magneto-oscillations with an amplitude that is orders of magnitude larger than the $B=0$ ratchet current recently detected in metamaterials of the same design and similar parameters~\cite{Hild2024,Yahniuk2024}.} We show that the oscillations are caused by the Shubnikov-de Haas effect. \bl{A specific feature of the magneto-ratchet oscillations is the nontrivial relation of their period and phase to those of the Shubnikov-de Haas oscillations (SdHO) of the longitudinal resistance.} We show that this is due to the fact that while the ratchet effect is generated locally, e.g., at the  boundaries of the triangles, the magnetotransport mostly reflects the areas outside the triangles which may have a different carrier density. A further enhancement of the ratchet current is observed at cyclotron resonance (CR). The theory of the CR enhanced ratchet currents shows that the resonances have a Fano shape defined by the 2D lateral asymmetry parameter. Strikingly, using  GHz radiation   we also detected the ratchet effect caused by electron spin resonance (ESR). We show that the resonant absorption causes a Seebeck ratchet signal, generated in the course of the photoinduced spatially-modulated heating of the free carriers accompanied by a periodic modulation of the equilibrium carrier density. The developed theory of the magneto-ratchet current describes well all features of the observed effects.

 	\section{Sample and methods}
 	\label{samples_methods}
 	
 	The sample design and experimental setup are sketched in Fig.\,\ref{fig1}. Figure\,\ref{fig1}(a) shows the sample cross section. \bl{The metamaterial is fabricated on a highly doped silicon wafer, acting as a global back gate, covered with a 285\,nm SiO$_2$ dielectric layer.} On top of this, a graphite film is deposited and structured with equilateral triangular holes (antidots) arranged in a square lattice with a period of 1000\,nm, see Fig.\,\ref{fig1}(b) and (c). The side length of the triangles is 600\,nm. The structured graphite film consisting of five graphene layers is used as a patterned gate. A monolayer of graphene encapsulated in 30\,nm thick hexagonal boron nitride (hBN) is deposited on the graphite patterned gate, see Fig.\,\ref{fig1}(b). Further details on sample preparation  can be found in Ref.\,\cite{Yahniuk2024}, where the same technique for the samples fabrication was used. Using reactive ion etching, the graphene is formed into a Hall bar configuration with ohmic chromium-gold contacts, see Fig.\,\ref{fig1}(c). The length and width of the sample~\#A (\#B) were 11\,$\mu$m (16\,$\mu$m) and 5\,$\mu$m (2.5\,$\mu$m), respectively. Fig.\,\ref{fig1}(b) shows the photograph of sample~\#A with red-bordered triangles highlighting the shape and position of a line of triangles across the Hall bar.  
 	The sample resistance $R_{xx}$ and the Hall resistance $R_{xy}$ were measured at a temperature of $T= 4.2$\,K. In the studied range of patterned gate ($|\Vp|\leq \SI{0.6}{V})$ and back gate voltages ($|\Vbg|\leq \SI{5}{V}$), the electron (hole) density was reaching $2.6 \times 10^{11}$\,cm$^{-2}$ for sample~\#A ($2.4 \times 10^{11}$\,cm$^{-2}$ for sample~\#B), with a corresponding carrier mobility of $44\times10^4$\,cm$^2$/Vs ($53 \times 10^4$\,cm$^2$/Vs) and momentum relaxation time 2.6\,ps (2.9\,ps). The transport characteristics are shown in Fig\,\ref{figA1} in App.\,\ref{appendixA}.  
 	
  		\begin{figure*}[t] 
 	\centering
 	\includegraphics[width=\linewidth]{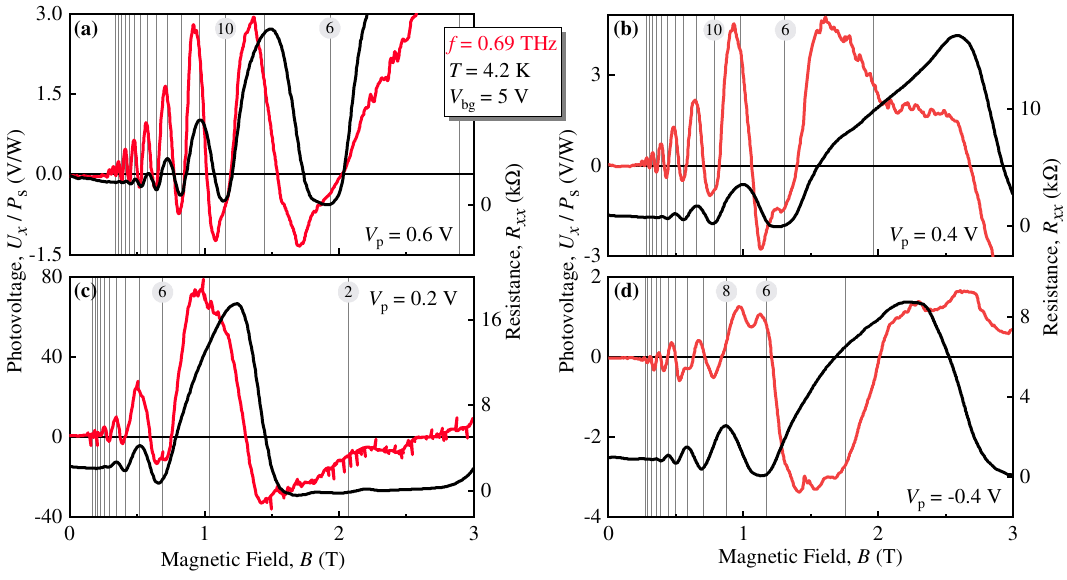}
 	\caption{Photovoltage measured as a function of magnetic field in $x$-direction (along the triangles' height), normalized to radiation power, for different patterned gate voltages $\Vp$ at a fixed gate voltage $\Vbg$. The data are obtained in sample \#A for frequency 0.69\,THz and radiation electric field vector parallel to $x$-direction ($\alpha =\ang{90}$). The corresponding resistance $R_{xx}$ is represented by the black traces on the right axes. The vertical lines indicate even filling factors of $R_{xx}$ counted from higher to lower magnetic fields, see numbers in circles.
 	}
 	\label{Fig2}
 \end{figure*}

 	To drive the ratchet currents, we used normal incident monochromatic and linearly polarized  radiation with frequencies ranging from several THz to tens of GHz. The corresponding wavelength $\lambda$ in the interval from about $100$\,$\mu$m to 6.6\,mm was much larger than the lattice period $d$ (1\,$\mu$m). 	
 	
 	The THz radiation was generated by a line-tunable molecular gas laser optically pumped by a cw CO$_2$ laser. In the experiments described below, three radiation frequencies 2.54\,THz ($\lambda =118\,\mu$m, photon energy $\hbar \omega = 10.5$\,meV), 1.62\,THz ($184\,\mu$m, 6.7\,meV), and 0.69\,THz ($432\,\mu$m, 2.9\,meV) were used. The laser lines were obtained with methanol, difluoromethane, and formic acid as active gases, respectively. The radiation was focused through two $z$-cut quartz windows of the cryostat by a gold-coated parabolic mirror. The laser beam profile at the sample position was measured with a pyroelectric camera\,\cite{Ganichev1999,Herrmann2016}. It had a Gaussian shape with a full width at half maximum of $\SI{1.5}{mm}$ (2.54\,THz), $\SI{1.8}{mm}$ (1.62\,THz), and 3\,mm (0.69\,THz). The corresponding beam areas were $A_\mathrm{beam} \approx 0.018$\,cm$^2$ for $2.54$\,THz, 0.025\,cm$^2$ for $1.62$\,THz, and 0.071\,cm$^2$ for $0.69$\,THz. The laser spot areas were much larger than the areas of the Hall bars $A_\mathrm{s}$ ensuring  uniform irradiation. The beam power was measured by a power meter and was about 40, 80, and 15\,mW for $f=2.54$, $f = 1.62$, and $f=0.69$\,THz, respectively. The power on the sample can be estimated as $P_\mathrm{s}= P \times A_\mathrm{s}/A_\mathrm{beam}$, with $P_\mathrm{s}(2.54$\,THz) = $2.9\times 10^{-4}$\,mW, $P_\mathrm{s}(1.62$\,THz) = $4.1\times 10^{-4}$\,mW, and $P_\mathrm{s}(0.69$\,THz) $= 0.28\times 10^{-4}$\,mW. Note that the estimate of $P_{\rm s}$ is given for sample~\#A with a transmission coefficient of about 0.7 per mm for the quartz windows (5\,mm total thickness).
 	
 	For generating radiation in the GHz regime, a Schottky diode with frequency multiplication was used. Four frequencies were obtained in the range from 45 to 75\,GHz with optical powers of about 150\,mW. The beam was focused on the sample through two  windows made out of polymethylpentene (TPX) and one cold diamond window with gold-coated parabolic mirrors. Each window had a thickness of 2\,mm. Taking into account the transmission coefficients of TPX ($\approx 0.95$ per mm) and poly-crystalline diamond ($\approx 0.71$ per mm) in this frequency range, one can estimate the radiation power reaching the sample, $P_\mathrm{s}(60$\,GHz)$ = 4.3\times 10^{-5}$\,mW. Note that we have neglected the influence of possible Fabry-P\'{e}rot interference which can strongly modify the radiation power.

	\begin{figure*}[t] 
	\centering
	\includegraphics[width=\linewidth]{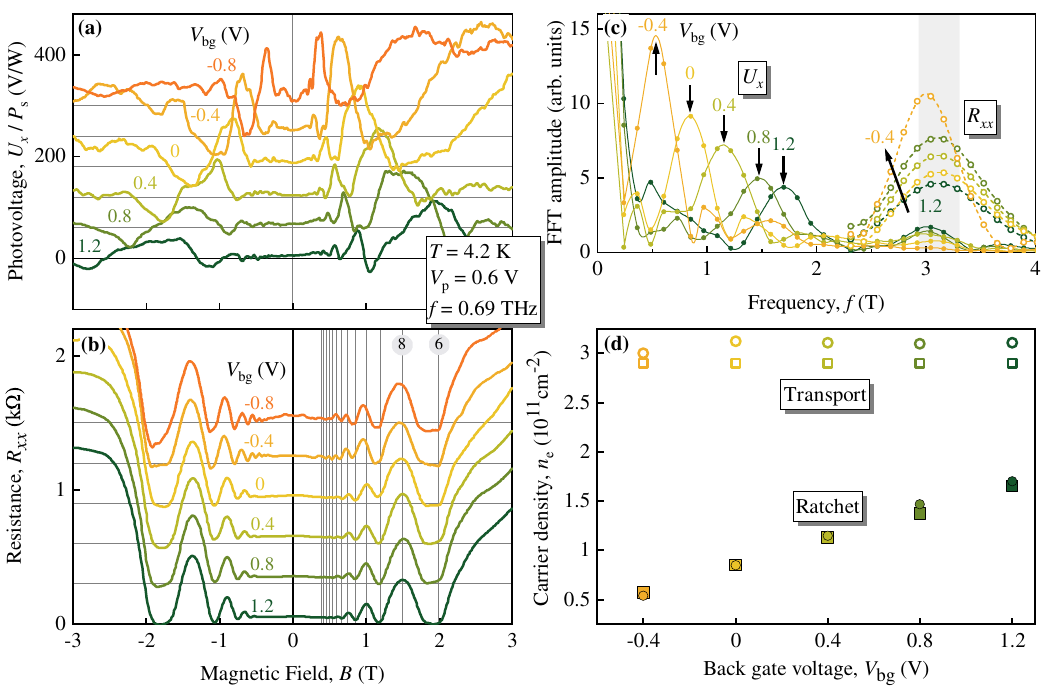}
	\caption{Panel (a): Evolution of the normalized photovoltage $U_x/P_\mathrm{s}$ with variation of the back gate voltage $\Vbg$ measured in the direction along the height of the triangles. The data were obtained for the azimuth angle $\alpha =\ang{90}$ and a patterned gate voltage $\Vp=0.6$\,V. The traces are shifted by 60\,V/W for clarity. Panel (b): The corresponding resistance $R_{xx}$ shifted by 0.3\,k$\Omega$. The vertical lines indicate even filling factors of $R_{xx}$ starting with 6 at high magnetic fields, see numbers in circles. Panel (c): the Fast Fourier transformation (FFT) analysis of the photosignal $U_x$ (full circles and solid lines) and the longitudinal resistance $R_{xx}$ (open circles and dashed lines) obtained for different back gate voltages indicated by numbers close to the curves. The spline lines are a guide for the eye. Panel (d): Back gate voltage dependence of the carrier density $n_{\rm e}$ obtained from the oscillations of the photosignal (full symbols) and $R_{xx}$ (open symbols). Circles show the densities obtained from the FFT, and rectangles show the densities obtained from the filling factor analysis. The data were obtained in sample~\#A.
	}
	\label{Fig3}
\end{figure*}

 	In most  experiments, the in-plane electric field vector $\bm{E}(\omega)$ of radiation was oriented parallel to the source-drain direction of the Hall bar ($x$-direction), i.e., along the height of the triangles. In some experiments, the electric field vector $\bm$ was rotated by  a crystal quartz lambda-half plate,  see  Fig.\,\ref{fig1}. Its orientation with respect to $x$-direction is defined by the azimuth angle $\alpha$.

 	 \bl{Measurements were made at liquid helium temperature $T=4.2$~K at THz frequencies and at $T=2$~K at GHz frequencies.} The warm windows of the cryostat were additionally covered with thin black polyethylene foils to prevent excitation by visible and near-infrared light. The photocurrent measurements were performed in the Faraday geometry with a magnetic field $B$ up to $\pm3$\,T applied perpendicular to the graphene layer plane, see Fig.\,\ref{fig1}(c). The photovoltage drop $U_x$ ($U_y$) was recorded from the source-drain contact pair 6-1 along the Hall bar and from the pair 5-7 perpendicular to it. The signals were measured  using the standard lock-in technique locked to the optical chopper frequency of 130\,Hz (THz laser radiation) or the Schottky diode electronics (GHz radiation).

 	\section{Results}
 	\label{results}

By illuminating the sample with linearly polarized radiation ($\bm E\parallel \hat{\bm x}$) of frequency 0.69\,THz while sweeping the applied magnetic field, we observed the photosignal oscillating with $B$. Figure\,\ref{Fig2} shows the photovoltage normalized to the radiation power, $\nUx$, for different  patterned gate voltages $\Vp$ at fixed global back gate voltage $\Vbg$. A photosignal with qualitatively the same behavior was found for all frequencies and gate voltage combinations used. Note that the $B=0$ photosignal is well detectable but has an amplitude that is orders of magnitude smaller than that in the presence of a magnetic field. It has been studied in detail in Ref.\,\cite{Yahniuk2024} and was shown to be caused by the linear ratchet effect.  \bl{The observation of the giant photosignal induced by the magnetic field directly indicates that it is caused by the magneto-ratchet effect. Indeed, even in graphene on a substrate that breaks the structural inversion symmetry, the generation of a direct current in the bulk under normal incidence is forbidden and indeed has not been observed; see a review in Ref.\,\cite{Glazov2014} and Secs.\,\ref{theory} and \ref{discussion}  for more details. This conclusion is further supported by the intensity dependencies, which show that the signal scales linearly with the radiation intensity and by the characteristic polarization dependencies, see Fig.\,\ref{figA2} and \ref{figA3} in  Appendix~\ref{appendixB}.}

 	\begin{figure*}[t] 
 		\centering
 		\includegraphics[width=\linewidth]{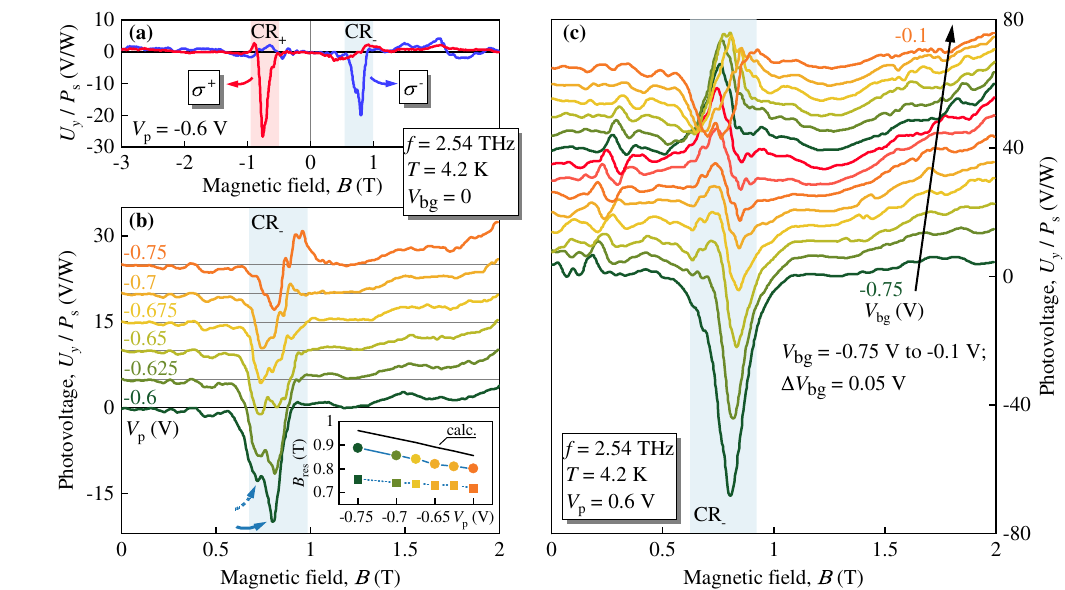}
 		\caption{Normalized photosignal $U_y/P_\mathrm{s}$ measured along the $y$-direction as a function of magnetic field for the azimuth angle $\alpha=0$ with $f=2.54$\,THz. Panel (a): Photosignal traces obtained for right-handed ($\sigma^+$, red curve) and left-handed ($\sigma^-$, blue curve) circularly polarized radiation with $\Vbg=0$. The emerging CR is highlighted by a correspondingly colored area and is named as CR$_+$ (CR$_-$) for $\sigma^+$ ($\sigma^-$). Panel (b): Photosignal traces obtained for left-handed circularly polarized radiation, $\Vbg=0$ and different patterned gate voltages. The solid and dashed blue arrow indicate the two observed resonances. The inset shows the magnetic field positions $B_{\rm res}$ of the resonances (colored circles) as a function of $\Vp$. The dashed (solid) blue arrow in the main panel corresponds to the circles connected by a dashed (solid) blue line. The black solid line was calculated using the carrier densities from transport characterization. Panel (c): Photosignal traces for $\Vp=0.6$\,V  and different back gate voltages varying from -0.75\,V to -0.1\,V with the step of 0.05\,V. The traces in panels (b) and (c) are up shifted by 5\,V/W for clarity. The data were obtained in sample~\#A.}
 		\label{Fig5}
 	\end{figure*}

 	To introduce the main features of the observed magneto-ratchet effect, we start with the results obtained at a frequency of 0.69\,THz. 
 	  First of all, as shown above, the ratchet current increases by orders of magnitude in the presence of a magnetic field. In addition, the magneto-ratchet signal is sign-alternating and has magneto-oscillations that grow exponentially in the low-$B$ region. Figures\,\ref{Fig2}(a)--(d) show the magnetic field dependence of the ratchet signal together with the longitudinal resistance $R_{xx}$ measured under the same conditions in the absence of illumination. A comparison with the magneto-resistance traces, showing  conventional SdHO, demonstrates that, for certain combinations of gate voltages, the period and phase of the magneto-oscillations in the ratchet current follow that of the SdHO. Indeed, in Fig.\,\ref{Fig2}(a)--(c) the minima and maxima of the ratchet oscillations exactly follow those of the SdHO in the resistance, reproducing the evolution of the latter with changes of patterned gate voltage $\Vp$. These observations provide a clear indication that the magneto-oscillations in both the ratchet current and resistance have a common nature and originate from Landau quantization. The results for sample~\#B are shown in the Appendix~\ref{appendixC}.

 		\begin{figure*}[t]
 		\centering
 		\includegraphics[width=\linewidth]{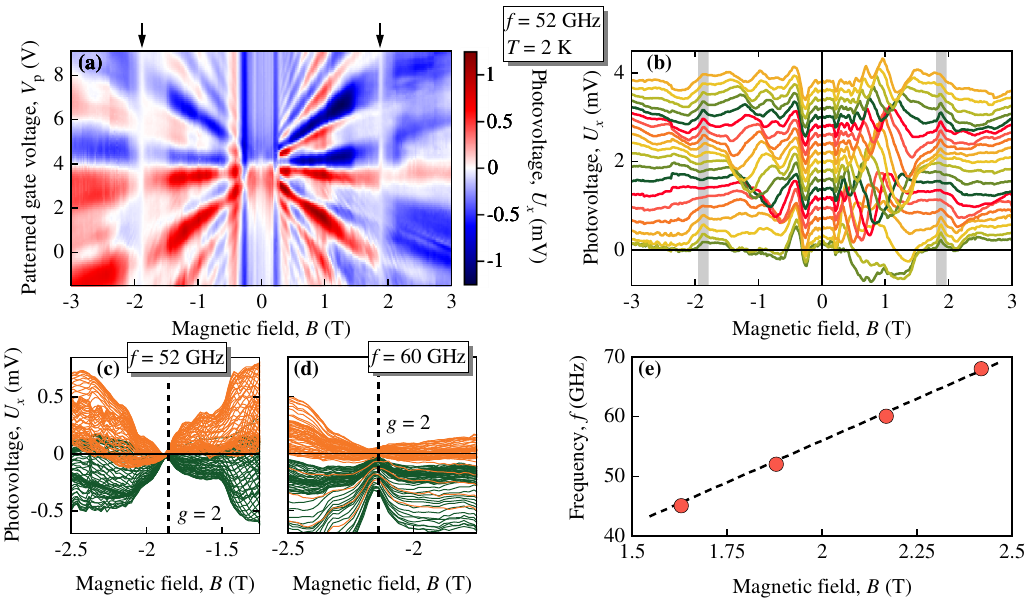}
 		\caption{Magnetic field dependencies of the photocurrent excited by the microwave radiation along the Hall bar. The data are obtained for zero back gate voltage and the azimuth angle $\alpha =\ang{90}$.  Panels (a)--(c) show the photovoltage in response to the radiation with $f=52$\,GHz and different patterned gate voltages. The photovoltage amplitudes in the panel (a) are indicated by the color bar on the right. The traces in the panel (b) show the magneto-ratchet signal for patterned gate voltages varied from $\Vp = 0.03$ to 0.6\,V with a step of 0.03\,V. The traces are  shifted vertically by 0.2\,mV for clarity. Vertical arrows in the color map, panel (a), and gray shaded regions in panel (b) indicate the position of the ESR that results in a vanishing photovoltage. Panels (c) and (d) show a zoom of the data obtained for $f=52$ and 60\,GHz. The traces were recorded for different patterned gate voltages varying from $\Vp = -0.6$ to 0.6\,V with a step of 0.01\,V. The positive (negative) $\Vp$ are shown in orange (green) color. They are shown without a vertical shift. The vertical dashed lines show the positions of the ESR for the Land\'e factor $g=2$. Panel (e) shows the evolution of the magnetic field position of the ESR with the excitation frequency. The dashed line is a calculated line for $g=2$. All data were recorded in sample~\#A.}
 		\label{Fig7}
 	\end{figure*}

 	At the same time, different choices of $\Vp$ and $\Vbg$ show that the magneto-oscillations in the ratchet current can become much more complex and may not follow that of the SdHO in $R_{xx}$, see Fig.\,\ref{Fig2}(d). To study these differences in detail, we varied the back gate voltage from 1.2\,V to $-$0.8\,V at a fixed  patterned gate voltage  $\Vp=0.6$\,V and compared the resulting traces with the transport data. The result is shown in Fig.\,\ref{Fig3}. It shows that the magneto-oscillations in the ratchet current vary significantly with the back gate voltage, see Fig.\,\ref{Fig3}(a), while the SdHO in the resistance, Fig.\,\ref{Fig3}(b), remain unaffected. Similar behavior was  also observed for the photosignal $\nUy$ measured across the Hall bar, i.e. along the bases of the triangles; see Fig.\,\ref{FigA4} in the Appendix~\ref{appendixB}.
 	
 	The analysis of these data, presented in Sec.~\ref{discussion} below, allows us to relate the distinct form of oscillations in the ratchet and the resistance to different spatial regions of current generation. Namely, the ratchet current arises due to the reduced symmetry of the metamaterial, and is thus  sensitive to both $\Vp$ and $\Vbg$, which define the THz near field and the scalar potential at the positions of the antidots. In contrast, SdHO in $R_{xx}$ are also present in uniform samples, and  in our case are dominated by large regions of uniform electron density between the antidots, making them almost insensitive to variations in $\Vbg$.
 	
 	We now turn to the  results obtained at the much higher frequency of 2.54\,THz, where, in addition to magneto-oscillations, a resonant enhancement of the magneto-ratchet current, associated with CR, was observed, see Fig.\,\ref{Fig5}. To conclude on the origin of the observed resonant enhancement of the ratchet current, we performed measurements using circularly polarized radiation. The results shown in Fig.\,\ref{Fig5}(a) clearly demonstrate that the resonant response to the right (left) circularly polarized radiation is detected for one polarity of magnetic field (labeled CR$_+$ and CR$_-$, respectively), which is in full agreement with the well-established Drude theory. For certain combinations of the back and patterned gate voltages two resonances are clearly resolved, as illustrated by the blue solid and dashed arrows in Fig.\,\ref{Fig5}(b). Figure~\ref{Fig5}(b) shows the evolution of the observed resonances as the patterned gate voltage is varied. Their corresponding magnetic field positions are shown  in Fig.\,\ref{Fig5}(e) as a function of the applied patterned gate voltage. It can be seen that, as expected for CR in graphene, the resonance positions shift to lower magnetic field upon decreasing of the gate voltage, which decreases the carrier density. For comparison, the solid line represents the calculated positions of CR with the carrier density estimated from transport measurements. The shape and position of the resonance is also affected by the variation of the back gate voltage. This is shown in Fig.\,\ref{Fig5}(c). An analysis of this complex behavior is presented below in the Sec.\,\ref{discussion}.

 	Finally, we present the results obtained with radiation at much lower frequencies in the tens of gigahertz range. Figure~\ref{Fig7}(a) and (b) shows the magnetic field dependence of the ratchet current excited by the radiation at $f=52$\,GHz. The data are obtained for zero back gate voltage and for different values of the patterned gate voltage. Strikingly, in addition to  magneto-oscillations similar to those detected at high frequency, this plot shows a new feature: a gate voltage-independent resonance at magnetic field $B = \pm 1.88$\,T, see downward arrows in panel (a) and dashed vertical lines in panel (b). At this field, the ratchet signal disappears. This can be seen even more clearly in Fig.\,\ref{Fig7}(a), which is a zoom of Fig.\,\ref{Fig7}(b), where the traces are not vertically shifted. Such a resonance feature was detected for all frequencies. Figure\,\ref{Fig7}(e) shows that its magnetic field position increases linearly with increasing frequency. Note that the vanishingly small signal for all patterned gate voltages was detected for $f=52$\,GHz only, see Fig.~\ref{Fig7}(c), for other frequencies we obtained that the signal decreases at resonance but does not necessarily approach zero, see Fig.~\ref{Fig7}(d) for $f=60$\,GHz. As we show below in Sec.\,\ref{discussion}, the resonance position corresponds to the electron spin resonance with Land\'e factor $g=2$, which is expected for graphene structures.

 	\section{Theory}
 	\label{theory}
 	The ratchet current is allowed in systems with broken space inversion symmetry. In the periodically modulated structures, e.g., metamaterials, there are two factors resulting in the ratchet effect. Namely, there is a noncentrosymmetric  static potential $V(\bm r)$ periodic with the 2D structure period, and the near field appearing as a result of diffraction on the triangle edges. The electric field acting on the carriers $\bm E(\bm r, t) = E_0(\bm r)\bm e \exp(-i\omega t)+c.c.$ has the near-field amplitude $E_0(\bm r)$ also periodic with the spatial period of the metamaterial ($\bm e$ is the polarization vector). 
 	The microscopic parameter which fixes the structure asymmetry is the 2D vector $\bm \Xi$ given by
 	\begin{equation}
 		\label{Xi}
 		\bm \Xi = \overline{E_0^2(\bm r) \bm \nabla V(\bm r)},
 	\end{equation}
 	where the overline means averaging over the spatial period of the structure.
 	
 	We begin with a discussion of ratchets with 1D modulation. This will allow us to derive the ratchet current in 2D metamaterials.
 	In Ref.\,\cite{Hubmann2020}, a theory of magneto-oscillations of the ratchet current was developed for the graphene-based structures with 1D modulation where the vector $\bm \Xi$, Eq.\,\eqref{Xi} had only one component. 
 	It was shown that the magnetoratchet current has two contributions. One is caused by the dynamic carrier-density redistribution (DCDR) taking place because of a simultaneous action of the static force $(-1/e) \bm \nabla V(\bm r)$ from the ratchet potential and the ac force $e\bm E(\bm r,t)$ of the radiation. The other mechanism of the ratchet current formation is caused by heating of carriers by radiation which is  inhomogeneous due to near-field modulation of the ac field amplitude. The inhomogeneous carrier temperature profile results in the so-called Seebeck ratchet current.
 	Both contributions were shown to have oscillating dependence on the magnetic field caused by the Landau quantization~\cite{Hubmann2020a,Budkin2016a,Moench2023b}. They are in phase with the Shubnikov-de~Haas oscillations in $R_{xx}$,
 	\begin{equation}
 		\label{sdho}
 		R_{xx}(B) = R_{xx}(0)(1+2\delta_c),
 	\end{equation}
 	and are described by the same factor $\delta_c$
 	\begin{equation}
 		\label{deltac}
 		\delta_c = 2\cos({\pi \varepsilon_{\rm F}\over\hbar\omega_c})\exp(-{\pi\over\omega_c\tau_q}){\chi\over\sinh \chi},
 	\end{equation}
 	where $\varepsilon_{\rm F}$ is the Fermi energy,  	$\omega_c=eBv_0^2/\varepsilon_{\rm F}$  is the cyclotron frequency, with $v_0$ being the Dirac fermion velocity in graphene, $\tau_q$ is the quantum relaxation time, and $\chi=2\pi^2 k_{\rm B}T/(\hbar\omega_c)$.  	It was shown that the oscillating ratchet current has an amplitude much larger than the ratchet current in the absence of the magnetic field due to the enhancement factor $(\varepsilon_{\rm F}/\hbar\omega_c)^2 \gg 1$. For the 1D modulation in the $x$ direction, when $\Xi_x \neq 0$, $\Xi_y =0$, the ratchet current components are given by the real and imaginary parts of the following expression
 	\begin{multline}
 		(j_x+ij_y)_{\rm 1D} = \Xi_x 
 		\qty({2\pi \varepsilon_{\rm F}\over\hbar\omega_c})^2 \delta_c
 		\\ \times
 		(C_0 + P_{\rm L1}C_{\rm L1}+ P_{\rm L2}C_{\rm L2}+ P_{\rm circ}C_{\rm circ}).
 	\end{multline}
 	Here $P_{\rm L1,L2,circ}$ are the Stokes parameters of the radiation ($P_{\rm L1,L2}$ are the linear polarization degrees and $P_{\rm circ}$ is the circular polarization), and $C_{\rm 0,L1,L2,circ}$ are functions of the frequency, transport relaxation time $\tau_{\rm tr}$, and magnetic field, see Appendix of Ref.\,\cite{Hubmann2020}.
 	
 	The result presented above contains a product of the asymmetry factor $\Xi_x$ and the function determined by characteristics of the graphene sample such as the Fermi energy and relaxation times. This multiplicative form allows one to generalize the results to the systems with 2D modulation. Indeed, in the linear-in-$\Xi_{x,y}$ regime one can consider 2D modulation as two superimposed structures with independent modulations in the $x$ and $y$ directions, and apply the 1D result to the components of the ratchet current along and perpendicular to the 1D modulation directions~\cite{Hild2024,Yahniuk2024}. This procedure yields
 	\begin{multline}
 		\label{j_2D}
 		j_x+ij_y = (\Xi_x+i\Xi_y) 
 		\qty({2\pi \varepsilon_{\rm F}\over\hbar\omega_c})^2 \delta_c
 		\\ \times
 		(C_0 + P_{\rm L1}C_{\rm L1}+ P_{\rm L2}C_{\rm L2}+ P_{\rm circ}C_{\rm circ})
 	\end{multline}
 	for the magneto-oscillations of the ratchet current in 2D metamaterial based on monolayer graphene, with the same functions $C_{\rm 0,L1,L2,circ}$ as for 1D modulation.
 	
 	In the vicinity of the CR, the ratchet current has resonances similar to the case of ratchets in systems with massive carriers considered in Ref.\,\cite{Moench2023b}. Expanding the coefficients $C_{\rm 0,L1,L2,circ}$ near the CR, we obtain
 	\begin{equation}
 		\label{jres}
 		j_{x,y}(\omega\approx \omega_c)= {e^3 v_0^2\over 8\pi\hbar^2 \varepsilon_{\rm F}\omega^3}\qty({2\pi \varepsilon_{\rm F}\over\hbar\omega_c})^2 \delta_c {\Phi_{x,y}(\epsilon)\over 1+\epsilon^2},
 	\end{equation}
 	where $\epsilon=(\omega-\omega_c)\tau_{\rm tr}$,
 	\begin{multline}
 		\label{Phi}
 		\Phi_x= (\epsilon\Xi_x-2\Xi_y)(1+P_{\rm L1})-(2\Xi_x+\epsilon\Xi_y)P_{\rm L2} 
 		\\
 		+ \qty[(1+\epsilon)\Xi_x+\epsilon\Xi_y]P_{\rm circ},
 	\end{multline}
 	and $\Phi_y$ is obtained from $\Phi_x$ via substitutions $\Xi_x \to \Xi_y$, $\Xi_y \to -\Xi_x$. The above expressions show that the $x$ and $y$ components of the polarization-dependent and independent ratchet currents near the CR have Fano shapes that are symmetric or asymmetric functions of the detuning $\epsilon$ depending on the ratio of the asymmetry parameters $\Xi_x$ and $\Xi_y$.	We emphasize that the magnetoratchet current Eqs.\,\eqref{j_2D}--\eqref{Phi} oscillates in magnetic field with the period defined by the carrier density in the area where the photocurrent is formed.
 	
 	
 	The ESR-induced ratchet current can be explained by the Seebeck ratchet mechanism caused by strong heating of the electron gas due to resonant radiation absorption~\cite{Bray2022}.
 	The inhomogeneous heating leads to the Seebeck ratchet effect  which is especially important in the resonant absorption conditions.	The corresponding contribution to the ratchet current $\bm j^{\rm S}$ is also governed by the asymmetry parameter $\bm \Xi$, Eq.\,\eqref{Xi}. Generalization of the results obtained for 1D modulated systems~\cite{Olbrich2009ratchet,Ivchenko2011,Nalitov2012,Olbrich2016,Budkin2016a} to the case of 2D modulation gives
 	\begin{equation}
 		\bm j^{\rm S} = -{1\over2e} {\partial \sigma\over \partial T}\, \overline{ T(\bm r) \bm \nabla V(\bm r)}.
 	\end{equation}
 	Here $\sigma(T)$ is the temperature-dependent conductivity of the system, which determines the radiation absorption. $ T(\bm r)$ is the carrier temperature which is higher than the lattice temperature due to radiation absorption and is periodically modulated because it is proportional to the near-field intensity. 
  
  In the ESR conditions, the carriers absorb the radiation very efficiently, and the spatial modulation of $ T(\bm r)$ has a high amplitude.	Due to the mechanism discussed above, this leads to a large ratchet current. The ESR-induced ratchet current varies with frequency because the maximum value of the absorption coefficient in the resonance is frequency dependent.

 	\section{Discussion}
 	\label{discussion}

 	The theory presented above explains our experimental findings presented in Sec.\,\ref{results}, including the origin of the ratchet effect in the 2D metamaterial, strong enhancement of the ratchet current in the presence of magnetic field, the emergence of magneto-oscillations of the ratchet current, the complex shape of the CR-enhanced magnetoratchet, and the observation of the ESR-induced ratchet. Below, we address these points in more detail. Altogether, they show that our results are well described by the developed theory of magneto-ratchet effect caused by the interplay of the THz near field of diffraction and the nonuniform electrostatic potential of the metamaterial controlled by the gate voltages.

 	\begin{figure}[t] 
 		\centering
 		\includegraphics[width=\linewidth]{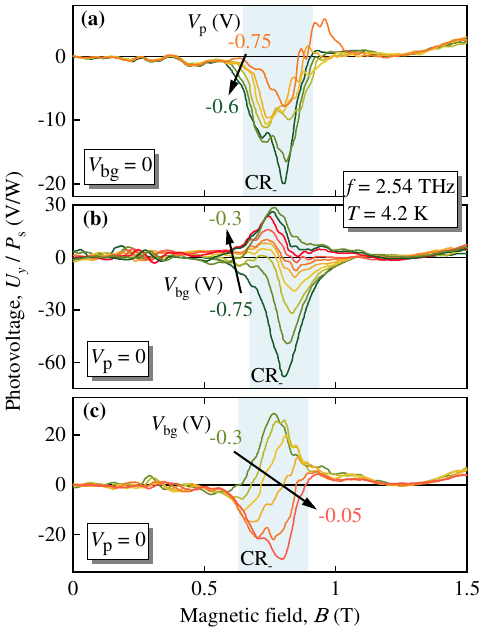}
 		\caption{Figure with identical data as presented in Fig.\,\ref{Fig5} but illustrated without a vertical offset. Panel (a) corresponds to the data in Fig.\,\ref{Fig5} (b), where $\Vp$ is varied $\Vbg$ is fixed, whereas panel (b) and (c) show the data from Fig.\,\ref{Fig5} (c) with different $\Vbg$ and fixed $\Vp$. The normalized photosignal was measured along the $y$-direction as a function of magnetic field for left-handed circularly polarized THz radiation ($\sigma^-$). The position of CR is highlighted by the blue shaded area and indicated with CR$_-$.}
 		\label{Fig5_1}
 	\end{figure}

 	To better understand the distinct role of patterned gate and back gate voltages ($\Vp$ and $\Vbg$) in our experiments, we first address the SdHO in the longitudinal resistance. Black traces in Fig.\,\ref{Fig2}(a)--(c) show that SdHO in $R_{xx}$ are highly sensitive to $\Vp$, while Fig.\,\ref{Fig3}(b) demonstrates their almost complete insensitivity to variation in $\Vbg$. This behavior is expected since the regions above the triangular holes (antidots) in the patterned gate occupy only 16 percent of the total graphene area. Outside the antidots, the density of carries is well fixed by the patterned gate that is about ten times closer to the graphene sheet than the back gate. Therefore, the period of SdHO in $R_{xx}$ is determined by $\Vp$ with a good accuracy, and SdHO can serve as probe of electron density $n_{\rm e}$ outside triangles. Expressing the cosine in Eq.\,\eqref{deltac} as $\cos(\pi\nu_{\rm f}/2)$ in terms of the filling factor,  $\nu_{\rm f}=2\pi\hbar n_{\rm e}/(e B)\equiv 2\varepsilon_F/\hbar\omega_c$, and fitting all minima and maxima of SdHO at even filling factors with the single parameter $n_{\rm e}$ (filling factor analysis), we find $n_{\rm e}=\{2.8,1.9,1\}\times10^{11}$\,cm$^{-2}$ for $\Vp=\{0.6,0.4,0.2\}$\,V in Fig.\,\ref{Fig2}(a)--(c), correspondingly, consistent with the density variation by $\Vp$ found from Hall measurements, see Appendix~\ref{appendixA}. The $B$-positions of even filling factors obtained in this way are shown as vertical lines in Figs.\,\ref{Fig2} and \ref{Fig3}.
 	
 	Now we turn back to the discussion of the ratchet effect. According to the theory presented in Sec.\,\ref{theory}, the observed photocurrent emerges due to broken inversion symmetry in the metamaterial, described by the 2D asymmetry parameter $\bm \Xi$, Eq.\,\eqref{Xi}. In the presence of magnetic field, the theory predicts the emergence of magneto-oscillations in the ratchet current, see Eq.\,\eqref{j_2D}. A giant enhancement of the ratchet current by the magnetic field, evident in all our observations, is due to the presence of the large factor $(2\pi\varepsilon_F/\hbar\omega_c)^2=(\pi\nu_f)^2$ in Eq.\,\eqref{j_2D}. This factor, absent for SdHO in $R_{xx}$, Eq.\,\eqref{sdho}, also explains why the ratchet magneto-oscillations can extend to much lower $B$ (higher $\nu_f$) than SdHO in $R_{xx}$, as can be clearly seen in Fig.\,\ref{Fig2}.
 	
 	In accord with our findings in Fig.\,\ref{Fig2}(a)--(c), in  case of weak and smooth modulation of electrostatic potential $V({\bm r})$ the theory predicts that magneto-oscillations in the ratchet current are in phase with the SdHO in $R_{xx}$, see Eq.\,\eqref{j_2D}. In general, however, the variation in $\Vp$ and $\Vbg$ in our experiments can lead to a strong modulation of electron density between the interior and exterior of the antidots, which results in a complex magneto-oscillation pattern of the ratchet signal, no longer following the behavior of $R_{xx}$. An instructive example is provided in Fig.\,\ref{Fig3}. Here, the variation in $\Vbg$ at fixed $\Vp$ results in significant changes of the ratchet signal, Fig.\,\ref{Fig3}(a), while the form of $R_{xx}(B)$ remains the same, in accord with the discussion above. The Fourier analysis  shows that the FFT spectrum of ratchet oscillations [solid lines in Fig.\,\ref{Fig3}(c)] includes two frequencies: the low-frequency peak in the FFT spectrum monotonously shifts with varying $\Vbg$, while the less pronounced high-frequency peak remains at the same position. FFT analysis of $R_{xx}(B)$ [dashed lines in Fig.\,\ref{Fig3}(c)] establishes that the high-frequency peak corresponds to the frequency of SdHO. It is natural to attribute these two frequencies to different densities (filling factors) inside and outside	triangular antidots. Indeed, the densities obtained from the low-frequency peak in the FFT spectrum of the ratchet current, see Fig.\,\ref{Fig3}(d), show  a linear increase with $\Vbg$, consistent with this interpretation.

 	This analysis demonstrates that in  case of strong modulation [note up to the factor of six difference in densities inside and outside antidots in Fig.\,\ref{Fig3}(d)] the ratchet current combines magneto-oscillations coming from variations in filling factors both inside and outside antidots, primarily controlled, correspondingly, by $\Vbg$ and $\Vp$. This reflects the nature of the ratchet current, emerging due to the interplay of asymmetric electrostatic field and near-field component of the THz field. These fields should be highly sensitive to variation in the density of states on both sides of triangle edges, giving rise to the observed complexity of the magnetoratchet signal. Summarizing, the distinct form of oscillations in the ratchet and resistance can be attributed to different spatial regions of the current formation. Namely, the ratchet current emerges due to the reduced symmetry of the metamaterial and thus is sensitive to both $\Vp$ and $\Vbg$ defining the near THz field and the static periodic potential at positions of antidots. By contrast, SdHO in $R_{xx}$ are present also in uniform samples and, in our case, are dominated by large areas with uniform electron density between the antidots, thus being almost insensitive to variations in $\Vbg$.
 	
 	Apart from giant oscillations, Eq.\,\eqref{j_2D} reveals that the magnetoratchet current scales as the second power of the THz field and has a specific polarization dependence determined by the Stokes parameters $P_{\rm L1,L2,circ}$. Both these characteristics have also been experimentally confirmed, see Figs.\,\ref{figA2} and \ref{figA3} in Appendix~\ref{appendixB}.
 	
 	We turn to discussion of the CR-enhancement of the ratchet current observed using circularly polarized radiation of $f=2.54$\,THz, see Fig.\,\ref{Fig5}. As mentioned in Sec.\,\ref{results}, a strong indication that this enhancement is related to the CR comes from its specific helicity dependence: the resonant ratchet current is observed only at positive (negative) $B$ for the left- (right-) handed radiation, see Fig.\,\ref{Fig5}(a), in agreement with the well established theory of the CR. An additional support comes from position of the resonance which shifts to higher $B$ with increasing $|\Vp|$ as illustrated in the inset to Fig.\,\ref{Fig5}(b). This behavior is consistent with the growth of the carrier concentration $n_{\rm e}$ in regions of the metamaterial outside the antidots which can be estimated from the transport data; the correspondent calculated position of the CR in the uniform monolayer of graphene, $B_{\rm res}=\hbar\omega\sqrt{\pi n}/(e v_0)$, is shown as black line labeled "calc." in the inset. While the density scaling is clearly the same, the calculated  $B_{\rm res}$ is slightly larger. This can be attributed to the different regions of the metamaterial where the ratchet current and SdHO in $R_{xx}$ are formed, as discussed above. 
 	
 	We now discuss the complex shape and fine structure of the observed CR in 2D ratchet current. In Fig.\,\ref{Fig5}(a)	as well as in few lower traces in Fig.\,\ref{Fig5}(b) the CR shape can be interpreted as combination of two distinct minima of usual Lorentzian form. Similar behavior was recently observed in a graphene metamaterial with 1D modulation \cite{Moench2023b}, in which it was shown that the two minima arise due to excitation of magnetoplasmon and cyclotron modes that can be observed simultaneously in the ratchet current. In the 2D metamaterial, the observed behavior of the CR is far more complex, as illustrated by strong variations of the shape and structure of the CR ratchet signal with $\Vp$ and $\Vbg$, Fig.\,\ref{Fig5}(b) and (c). To present these variations in more detail, we replot the data without vertical shift of individual traces in Fig.\,\ref{Fig5_1}. It can be seen that in the studied range of $\Vp$ and $\Vbg$, the ratchet oscillations outside the CR region are insensitive to $\Vp$ but are strongly sensitive to changes of $\Vbg$. In the CR region, sensitive to both gate voltages, one observes additional weak short-period oscillations, which may be responsible for the fine structure of CR including distinct two minima observed at certain combinations of $\Vp$ and $\Vbg$. The nature of these short-period oscillations is currently unclear, and can, e.g., stem from excitation of magnetoplasmon modes in the antidots.
  
 	More remarkable are changes in the overall shape of the CR, which may form a single Lorentzian peak/dip depending on gate voltages, or combine neighbouring peak and dip. This behavior is consistent with the theory presented in Sec.\,\ref{theory}, which predicts that the CR shape in 2D ratchet contains both symmetric (Lorentz) and antisymmetric (Fano) components. According to Eq.\,\eqref{jres}, their relative magnitude is controlled by the relation of the asymmetry parameters $\Xi_x$ and $\Xi_y$, which can vary with $\Vp$ and $\Vbg$. Taking into account that CR in the ratchet current essentially represents a resonant enhancement of magneto-oscillations and thus may have arbitrary sign depending om $\Vp$ and $\Vbg$, this explains on a qualitative level the complex evolution of the CR shape presented in Figs.\,\ref{Fig5} and \ref{Fig5_1}. 
 	
 	We finally discuss the observation of the features attributed to electron spin resonance in experiments using GHz radiation, see Fig.\,\ref{Fig7}. Indeed, the ESR condition $\hbar \omega= g \mu_\text{B} B$,  where $\mu_\text{B}= e\hbar/2 m_e$, for Land\'e factor $g=2$ expected for graphene gives $B_{\rm ESR}=m_e\omega/e$, which is exactly the position of the features observed for all frequencies $\omega/2\pi=\{45,52,60,68\}$\,GHz in the ratchet current in our experiments, see dashed line in the inset in Fig.\,\ref{Fig7}(a). According to Sec.\,\ref{theory}, the ESR contribution to the ratchet current occurs due to resonant direct optical transitions between spin subbands, resulting in the enhanced heating and corresponding Seebeck ratchet effect. This contribution competes with the ratchet mechanisms associated with indirect optical transitions (Drude absorption). Our results obtained at $\omega/2\pi=52$\,GHz demonstrate that the ESR and Drude contributions occasionally compensate each other  at this particular frequency, see Fig.\,\ref{Fig7}(c). At other frequencies, such complete compensation does not take place. This is caused by different frequency dependences of the Drude absorption and resonant absorption associated with the direct transitions, leading to ESR. 

 	\section{Summary}
 	\label{summary}
 	
 	Our results show that excitation of a 2D graphene-based metamaterial by THz/GHz radiation causes the magneto-ratchet current. The current exhibits $1/B$-periodic oscillations with an amplitude that is orders of magnitude larger than the ratchet current at zero magnetic field. While the oscillations are related to the SdH effect, they do not necessarily follow the SdH oscillations in the dc resistance. \bl{The distinct form of oscillations in ratchet signal and resistance confirms different spatial regions of the current formation. Indeed, as the ratchet current emerges due to the reduced symmetry of the metamaterial, it should be sensitive to both gate potentials defining the near field and the static periodic potential at positions of antidots, in agreement with our observations.}
  
    A further enhancement of the magneto-ratchet effect is observed in the region of the magnetic field where the cyclotron frequencies match the THz radiation frequency. The theory of the cyclotron resonance-induced ratchet effect describes the experimental findings well. In particular, it shows that the polarization-dependent and -independent magneto-ratchet currents near the CR have Fano resonance shapes that are symmetric or asymmetric functions of the detuning $\epsilon$, depending on the ratio of the lateral asymmetry parameters $\Xi_x$ and $\Xi_y$. 
  
  In addition to the CR-induced THz ratchet effect, we observed an different resonance response induced by GHz radiation. We show that this effect is caused by the electron spin resonance-induced ratchet current, which has the opposite sign to the magneto-ratchet current caused by the Drude absorption and DCDR mechanism. This leads to a decrease in the current magnitude at the resonance and, for some frequencies, to a complete cancellation of the ratchet current. The theory developed shows that this resonant response is caused by the Seebeck ratchet current resulting from the inhomogeneous heating due to the spatially modulated near-field effect. Resonant absorption increases the heating and gives rise to the resonant ratchet current. 
  
  Our results show that THz/GHz ratchet currents provide a novel experimental approach to study 2D metamaterials and fundamental properties of graphene layers.

 	\section{Acknowledgments}
 	\label{acknow}
 	 We acknowledge the financial support of the Deutsche Forschungsgemeinschaft (DFG, German Research Foundation) via Project-ID 314695032 – SFB 1277 (Subprojects A01,  A04, and A09) and via grant DM~1/6-1 (I.A.D.), DFG Project ID No. 426094608 (ER 612/2-1),
 	  and of the Volkswagen Stiftung Program (97738). S.D.G. is also grateful for the support of the European Union (ERC-ADVANCED “TERAPLASM,” Project No. 101053716). We also acknowledge the French Agence Nationale pour la Recherche (DeMeGRaS - ANR-19-GRF1-0006), the France 2030 program for Equipex+ Hybat project (ANR-21-ESRE-0026), and the Physics Institute of CNRS for Tremplin 2024 - Step - project.
 	   J.G.-S. acknowledge support from DFG GU 2528/1-1 695298. K.W. and T.T. acknowledge support from the JSPS KAKENHI (Grants No. 20H00354 and No. 23H02052) and World Premier International Research Center Initiative (WPI), MEXT, Japan. 
 	
 	\appendix
 	\counterwithin{figure}{section}
 	\setcounter{figure}{0}
 	
 	
 	 	\section{Transport characteristics}
 	\label{appendixA}
 	 			\begin{figure*}[t] 
 		\centering
 		\includegraphics[width=0.8\linewidth]{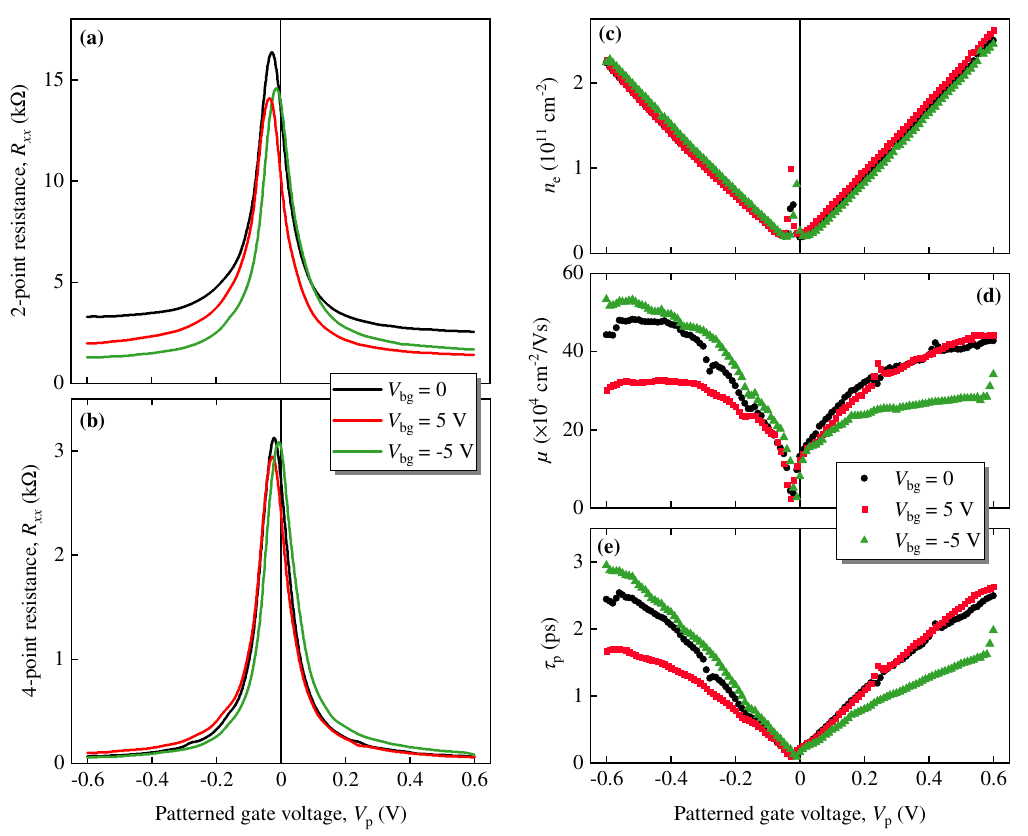}
 		\caption{Sample characteristics at $T=\SI{4.2}{K}$ as a function of patterned gate voltage for different back gate voltages, $\Vbg = \{0, \pm 5\,{\rm V}\}$: Panels (a) and (b) show the two-point and  four-point resistance $R_{xx}$ of sample~\#A, respectively, both measured along the height of triangles. Panel (c) shows the charge carrier density $n_\mathrm{e}$, (d) the carriers' mobility $\mu$, and (e) the momentum relaxation time $\tau_\mathrm{p}$.}
 		\label{figA1}
 	\end{figure*}
 	
 		We performed complementary electrical transport measurements in order to characterize the samples. Figure~\ref{figA1} shows the two-point [Fig.\,\ref{figA1}(a)] and four-point [Fig.\,\ref{figA1}(b)] resistance $R_{xx}$ of sample~\#A at $B=0$ as a function of the patterned gate voltage $\Vp$ within a range used in the optical experiments. We also measured the Hall resistance $R_{xy}$ at $B=\SI{0.2}{T}$ (not shown) to calculate different sample properties. In Fig.\,\ref{figA1}(c) the carrier density was determined as $n_\mathrm{e}= B/(e R_{xy})$, in Fig.\,\ref{figA1}(d) the mobility, given by $\mu=\left[n_\mathrm{e}e \rho_{xx}(B=0) \right]^{-1} $, and in Fig.\,\ref{figA1}(e) the momentum relaxation time, given by $\tau_p= \hbar\mu\sqrt{\pi n_\mathrm{e}}/(v_\mathrm{0}e)$, with $e$ being the electron charge, $v_\mathrm{0} = 10^6$\,m/s the Fermi velocity and $\rho_{xx} = W/(LR_{xx})$ the longitudinal resistivity, where $W$ and $L$ are the width and length of the measurement channel, respectively.

 	 	\section{Additional data obtained in sample \#A}
 	\label{appendixB}

 		\begin{figure}[t] 
 		\centering
 		\includegraphics[width=\linewidth]{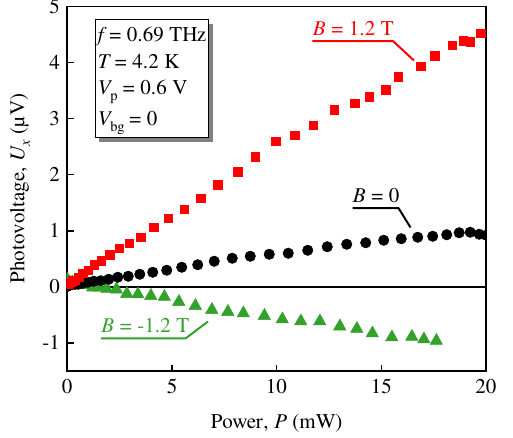}
 		\caption{Power dependence of the photovoltage measured in the $x$-direction in sample~\#A for three different values of the magnetic field, $B = \{0, \pm 1.2\,{\rm T}\}$.}
 		\label{figA2}
 	\end{figure}
 	
 	\begin{figure}[t] 
 		\centering
 		\includegraphics[width=\linewidth]{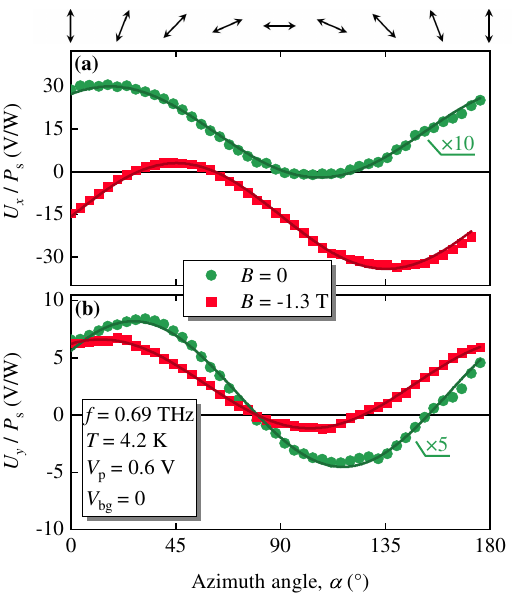}
 		\caption{Polarization dependence of the normalized photosignal of $U_x$ (a) and $U_y$ (b) obtained in sample~\#A at $B=0$\,T (green traces) and $B=-1.3$\,T (red traces).}
 		\label{figA3}
 	\end{figure}
 	
 		\begin{figure}[t] 
 		\centering
 		\includegraphics[width=\linewidth]{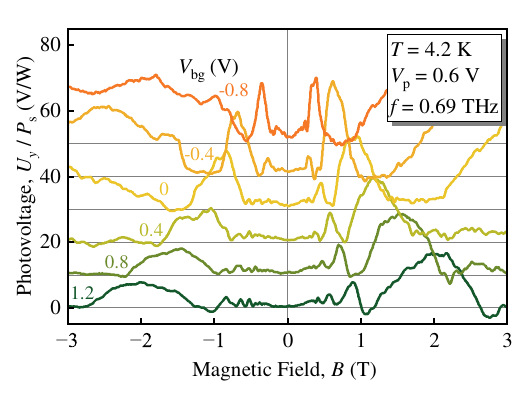}
 		\caption{Normalized photovoltage $\nUy$ measured along the basis of triangles in sample~\#A as a function of the magnetic field for various back gate voltages. The azimuth angle is $\alpha=\ang{90}$. The traces are up shifted by 10\,V/W for clarity.}
 		\label{FigA4}
 	\end{figure}
 	
 	Figure~\ref{figA2} shows the power dependence of the signal generated by the $0.69$\,THz radiation. The data shown were recorded at $B=0$ and $B=\pm\SI{1.2}{T}$, zero back gate voltage, and $\Vp=\SI{0.6}{V}$. It is seen that the photosignal scales linearly with the radiation power, i.e., as a square of the radiation electric field $E_0$, as expected for the linear ratchet effects, see Sec.~\ref{theory}. 
 	
 	Figure~\ref{figA3} shows the dependencies of the normalized photosignals $\nUx$ [Fig.\,\ref{figA3}(a)] and $\nUy$ [Fig.\,\ref{figA3}(b)] along the $x$- and $y$-direction, respectively. The patterned gate voltage was $\Vp=\SI{0.6}{V}$, the back gate voltage $\Vbg$ was set to zero, and the radiation frequency was $f=\SI{0.69}{THz}$. The obtained photosignal is a linear combination of Stokes parameters, see Eq.\,\ref{j_2D}, which for $B=0$ has been studied in detail in Refs.~\cite{Yahniuk2024,Hild2024}.
 	
  	Figure~\ref{FigA4} shows the photosignal $\nUy$ generated by the $0.69$\,THz radiation for several back gate voltages and fixed $\Vp=\SI{0.6}{V}$. Similar to Fig.~\ref{Fig3}(a), showing the corresponding $\nUx$ signal,  the radiation electric field $\bm E$ was parallel to the $x$-axis, i.e., oriented along the Hall bar.

 	\section{Results obtained in sample \#B}
 	\label{appendixC}
 	
 	\begin{figure}[t] 
 	\centering
 	\includegraphics[width=\linewidth]{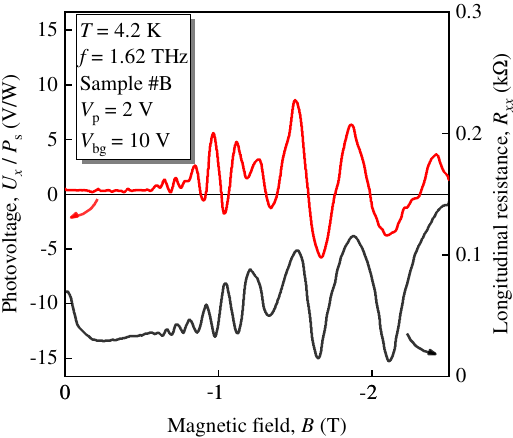}
 	\caption{Photovoltage $U_x$ (red curve) measured along the height of triangles in sample~\#B as a function of the magnetic field 
 		for the azimuth angle $\alpha=\ang{90}$. The black curve is the corresponding resistance.}
 	\label{FigA7}
 \end{figure}
 	
Red trace in Fig.~\ref{FigA7} shows the photovoltage $U_x$ generated in sample~\#B by the $1.62$\,THz radiation at $\Vp=\SI{2}{V}$ and $\Vbg=\SI{10}{V}$. The resistance measured in the same conditions is given by the black trace.


 	
 	%
 	%
 	%

 	\bibliography{all_lib}

\begin{thebibliography}{30}%
\makeatletter
\providecommand \@ifxundefined [1]{%
 \@ifx{#1\undefined}
}%
\providecommand \@ifnum [1]{%
 \ifnum #1\expandafter \@firstoftwo
 \else \expandafter \@secondoftwo
 \fi
}%
\providecommand \@ifx [1]{%
 \ifx #1\expandafter \@firstoftwo
 \else \expandafter \@secondoftwo
 \fi
}%
\providecommand \natexlab [1]{#1}%
\providecommand \enquote  [1]{``#1''}%
\providecommand \bibnamefont  [1]{#1}%
\providecommand \bibfnamefont [1]{#1}%
\providecommand \citenamefont [1]{#1}%
\providecommand \href@noop [0]{\@secondoftwo}%
\providecommand \href [0]{\begingroup \@sanitize@url \@href}%
\providecommand \@href[1]{\@@startlink{#1}\@@href}%
\providecommand \@@href[1]{\endgroup#1\@@endlink}%
\providecommand \@sanitize@url [0]{\catcode `\\12\catcode `\$12\catcode
  `\&12\catcode `\#12\catcode `\^12\catcode `\_12\catcode `\%12\relax}%
\providecommand \@@startlink[1]{}%
\providecommand \@@endlink[0]{}%
\providecommand \url  [0]{\begingroup\@sanitize@url \@url }%
\providecommand \@url [1]{\endgroup\@href {#1}{\urlprefix }}%
\providecommand \urlprefix  [0]{URL }%
\providecommand \Eprint [0]{\href }%
\providecommand \doibase [0]{https://doi.org/}%
\providecommand \selectlanguage [0]{\@gobble}%
\providecommand \bibinfo  [0]{\@secondoftwo}%
\providecommand \bibfield  [0]{\@secondoftwo}%
\providecommand \translation [1]{[#1]}%
\providecommand \BibitemOpen [0]{}%
\providecommand \bibitemStop [0]{}%
\providecommand \bibitemNoStop [0]{.\EOS\space}%
\providecommand \EOS [0]{\spacefactor3000\relax}%
\providecommand \BibitemShut  [1]{\csname bibitem#1\endcsname}%
\let\auto@bib@innerbib\@empty
\bibitem [{\citenamefont {Caldwell}\ \emph {et~al.}(2015)\citenamefont
  {Caldwell}, \citenamefont {Lindsay}, \citenamefont {Giannini}, \citenamefont
  {Vurgaftman}, \citenamefont {Reinecke}, \citenamefont {Maier},\ and\
  \citenamefont {Glembocki}}]{Caldwell2015}%
  \BibitemOpen
  \bibfield  {author} {\bibinfo {author} {\bibfnamefont {J.~D.}\ \bibnamefont
  {Caldwell}}, \bibinfo {author} {\bibfnamefont {L.}~\bibnamefont {Lindsay}},
  \bibinfo {author} {\bibfnamefont {V.}~\bibnamefont {Giannini}}, \bibinfo
  {author} {\bibfnamefont {I.}~\bibnamefont {Vurgaftman}}, \bibinfo {author}
  {\bibfnamefont {T.~L.}\ \bibnamefont {Reinecke}}, \bibinfo {author}
  {\bibfnamefont {S.~A.}\ \bibnamefont {Maier}},\ and\ \bibinfo {author}
  {\bibfnamefont {O.~J.}\ \bibnamefont {Glembocki}},\ }\bibfield  {title}
  {\bibinfo {title} {Low-loss, infrared and terahertz nanophotonics using
  surface phonon polaritons},\ }\href
  {https://doi.org/10.1515/nanoph-2014-0003} {\bibfield  {journal} {\bibinfo
  {journal} {Nanophotonics}\ }\textbf {\bibinfo {volume} {4}},\ \bibinfo
  {pages} {44} (\bibinfo {year} {2015})}\BibitemShut {NoStop}%
\bibitem [{\citenamefont {Xu}\ \emph {et~al.}(2017)\citenamefont {Xu},
  \citenamefont {Xie},\ and\ \citenamefont {Ying}}]{Xu2017b}%
  \BibitemOpen
  \bibfield  {author} {\bibinfo {author} {\bibfnamefont {W.}~\bibnamefont
  {Xu}}, \bibinfo {author} {\bibfnamefont {L.}~\bibnamefont {Xie}},\ and\
  \bibinfo {author} {\bibfnamefont {Y.}~\bibnamefont {Ying}},\ }\bibfield
  {title} {\bibinfo {title} {Mechanisms and applications of terahertz
  metamaterial sensing: a review},\ }\href {https://doi.org/10.1039/c7nr03824k}
  {\bibfield  {journal} {\bibinfo  {journal} {Nanoscale}\ }\textbf {\bibinfo
  {volume} {9}},\ \bibinfo {pages} {13864} (\bibinfo {year}
  {2017})}\BibitemShut {NoStop}%
\bibitem [{\citenamefont {Nemati}\ \emph {et~al.}(2018)\citenamefont {Nemati},
  \citenamefont {Wang}, \citenamefont {Hong},\ and\ \citenamefont
  {Teng}}]{Nemati2018}%
  \BibitemOpen
  \bibfield  {author} {\bibinfo {author} {\bibfnamefont {A.}~\bibnamefont
  {Nemati}}, \bibinfo {author} {\bibfnamefont {Q.}~\bibnamefont {Wang}},
  \bibinfo {author} {\bibfnamefont {M.}~\bibnamefont {Hong}},\ and\ \bibinfo
  {author} {\bibfnamefont {J.}~\bibnamefont {Teng}},\ }\bibfield  {title}
  {\bibinfo {title} {Tunable and reconfigurable metasurfaces and metadevices},\
  }\href {https://doi.org/10.29026/oea.2018.180009} {\bibfield  {journal}
  {\bibinfo  {journal} {Opto-Electronic Advances}\ }\textbf {\bibinfo {volume}
  {1}},\ \bibinfo {pages} {18000901} (\bibinfo {year} {2018})}\BibitemShut
  {NoStop}%
\bibitem [{\citenamefont {Yu}\ \emph {et~al.}(2018)\citenamefont {Yu},
  \citenamefont {Besteiro}, \citenamefont {Huang}, \citenamefont {Wu},
  \citenamefont {Fu}, \citenamefont {Tan}, \citenamefont {Jagadish},
  \citenamefont {Wiederrecht}, \citenamefont {Govorov},\ and\ \citenamefont
  {Wang}}]{Yu2018b}%
  \BibitemOpen
  \bibfield  {author} {\bibinfo {author} {\bibfnamefont {P.}~\bibnamefont
  {Yu}}, \bibinfo {author} {\bibfnamefont {L.~V.}\ \bibnamefont {Besteiro}},
  \bibinfo {author} {\bibfnamefont {Y.}~\bibnamefont {Huang}}, \bibinfo
  {author} {\bibfnamefont {J.}~\bibnamefont {Wu}}, \bibinfo {author}
  {\bibfnamefont {L.}~\bibnamefont {Fu}}, \bibinfo {author} {\bibfnamefont
  {H.~H.}\ \bibnamefont {Tan}}, \bibinfo {author} {\bibfnamefont
  {C.}~\bibnamefont {Jagadish}}, \bibinfo {author} {\bibfnamefont {G.~P.}\
  \bibnamefont {Wiederrecht}}, \bibinfo {author} {\bibfnamefont {A.~O.}\
  \bibnamefont {Govorov}},\ and\ \bibinfo {author} {\bibfnamefont
  {Z.}~\bibnamefont {Wang}},\ }\bibfield  {title} {\bibinfo {title} {Broadband
  metamaterial absorbers},\ }\href {https://doi.org/10.1002/adom.201800995}
  {\bibfield  {journal} {\bibinfo  {journal} {Adv. Opt. Mater.}\ }\textbf
  {\bibinfo {volume} {7}},\ \bibinfo {pages} {1800995} (\bibinfo {year}
  {2018})}\BibitemShut {NoStop}%
\bibitem [{\citenamefont {He}\ \emph {et~al.}(2019)\citenamefont {He},
  \citenamefont {Sun},\ and\ \citenamefont {Zhou}}]{He2019}%
  \BibitemOpen
  \bibfield  {author} {\bibinfo {author} {\bibfnamefont {Q.}~\bibnamefont
  {He}}, \bibinfo {author} {\bibfnamefont {S.}~\bibnamefont {Sun}},\ and\
  \bibinfo {author} {\bibfnamefont {L.}~\bibnamefont {Zhou}},\ }\bibfield
  {title} {\bibinfo {title} {Tunable/reconfigurable metasurfaces: Physics and
  applications},\ }\href {https://doi.org/10.34133/2019/1849272} {\bibfield
  {journal} {\bibinfo  {journal} {Research}\ }\textbf {\bibinfo {volume}
  {2019}},\ \bibinfo {pages} {1849272} (\bibinfo {year} {2019})}\BibitemShut
  {NoStop}%
\bibitem [{\citenamefont {Xiao}\ \emph {et~al.}(2020)\citenamefont {Xiao},
  \citenamefont {Wang}, \citenamefont {Liu}, \citenamefont {Zhou},
  \citenamefont {Jiang},\ and\ \citenamefont {Zhang}}]{Xiao2020}%
  \BibitemOpen
  \bibfield  {author} {\bibinfo {author} {\bibfnamefont {S.}~\bibnamefont
  {Xiao}}, \bibinfo {author} {\bibfnamefont {T.}~\bibnamefont {Wang}}, \bibinfo
  {author} {\bibfnamefont {T.}~\bibnamefont {Liu}}, \bibinfo {author}
  {\bibfnamefont {C.}~\bibnamefont {Zhou}}, \bibinfo {author} {\bibfnamefont
  {X.}~\bibnamefont {Jiang}},\ and\ \bibinfo {author} {\bibfnamefont
  {J.}~\bibnamefont {Zhang}},\ }\bibfield  {title} {\bibinfo {title} {Active
  metamaterials and metadevices: a review},\ }\href
  {https://doi.org/10.1088/1361-6463/abaced} {\bibfield  {journal} {\bibinfo
  {journal} {J. Phys. D: Appl. Phys.}\ }\textbf {\bibinfo {volume} {53}},\
  \bibinfo {pages} {503002} (\bibinfo {year} {2020})}\BibitemShut {NoStop}%
\bibitem [{\citenamefont {Xu}\ \emph {et~al.}(2022)\citenamefont {Xu},
  \citenamefont {Ren}, \citenamefont {Wei},\ and\ \citenamefont
  {Lee}}]{Xu2022}%
  \BibitemOpen
  \bibfield  {author} {\bibinfo {author} {\bibfnamefont {C.}~\bibnamefont
  {Xu}}, \bibinfo {author} {\bibfnamefont {Z.}~\bibnamefont {Ren}}, \bibinfo
  {author} {\bibfnamefont {J.}~\bibnamefont {Wei}},\ and\ \bibinfo {author}
  {\bibfnamefont {C.}~\bibnamefont {Lee}},\ }\bibfield  {title} {\bibinfo
  {title} {Reconfigurable terahertz metamaterials: From fundamental principles
  to advanced 6g applications},\ }\href
  {https://doi.org/10.1016/j.isci.2022.103799} {\bibfield  {journal} {\bibinfo
  {journal} {iScience}\ }\textbf {\bibinfo {volume} {25}},\ \bibinfo {pages}
  {103799} (\bibinfo {year} {2022})}\BibitemShut {NoStop}%
\bibitem [{\citenamefont {Olbrich}\ \emph {et~al.}(2016)\citenamefont
  {Olbrich}, \citenamefont {Kamann}, \citenamefont {K{\"o}nig}, \citenamefont
  {Munzert}, \citenamefont {Tutsch}, \citenamefont {Eroms}, \citenamefont
  {Weiss}, \citenamefont {Liu}, \citenamefont {Golub}, \citenamefont
  {Ivchenko}, \citenamefont {Popov}, \citenamefont {Fateev}, \citenamefont
  {Mashinsky}, \citenamefont {Fromm}, \citenamefont {Seyller},\ and\
  \citenamefont {Ganichev}}]{Olbrich2016}%
  \BibitemOpen
  \bibfield  {author} {\bibinfo {author} {\bibfnamefont {P.}~\bibnamefont
  {Olbrich}}, \bibinfo {author} {\bibfnamefont {J.}~\bibnamefont {Kamann}},
  \bibinfo {author} {\bibfnamefont {M.}~\bibnamefont {K{\"o}nig}}, \bibinfo
  {author} {\bibfnamefont {J.}~\bibnamefont {Munzert}}, \bibinfo {author}
  {\bibfnamefont {L.}~\bibnamefont {Tutsch}}, \bibinfo {author} {\bibfnamefont
  {J.}~\bibnamefont {Eroms}}, \bibinfo {author} {\bibfnamefont
  {D.}~\bibnamefont {Weiss}}, \bibinfo {author} {\bibfnamefont {M.-H.}\
  \bibnamefont {Liu}}, \bibinfo {author} {\bibfnamefont {L.~E.}\ \bibnamefont
  {Golub}}, \bibinfo {author} {\bibfnamefont {E.~L.}\ \bibnamefont {Ivchenko}},
  \bibinfo {author} {\bibfnamefont {V.~V.}\ \bibnamefont {Popov}}, \bibinfo
  {author} {\bibfnamefont {D.~V.}\ \bibnamefont {Fateev}}, \bibinfo {author}
  {\bibfnamefont {K.~V.}\ \bibnamefont {Mashinsky}}, \bibinfo {author}
  {\bibfnamefont {F.}~\bibnamefont {Fromm}}, \bibinfo {author} {\bibfnamefont
  {T.}~\bibnamefont {Seyller}},\ and\ \bibinfo {author} {\bibfnamefont {S.~D.}\
  \bibnamefont {Ganichev}},\ }\bibfield  {title} {\bibinfo {title} {Terahertz
  ratchet effects in graphene with a lateral superlattice},\ }\href
  {https://doi.org/10.1103/physrevb.93.075422} {\bibfield  {journal} {\bibinfo
  {journal} {Phys. Rev. B}\ }\textbf {\bibinfo {volume} {93}},\ \bibinfo
  {pages} {075422} (\bibinfo {year} {2016})}\BibitemShut {NoStop}%
\bibitem [{\citenamefont {Ganichev}\ \emph {et~al.}(2017)\citenamefont
  {Ganichev}, \citenamefont {Weiss},\ and\ \citenamefont
  {Eroms}}]{Ganichev2017}%
  \BibitemOpen
  \bibfield  {author} {\bibinfo {author} {\bibfnamefont {S.~D.}\ \bibnamefont
  {Ganichev}}, \bibinfo {author} {\bibfnamefont {D.}~\bibnamefont {Weiss}},\
  and\ \bibinfo {author} {\bibfnamefont {J.}~\bibnamefont {Eroms}},\ }\bibfield
   {title} {\bibinfo {title} {Terahertz electric field driven electric currents
  and ratchet effects in graphene},\ }\href
  {https://doi.org/10.1002/andp.201600406} {\bibfield  {journal} {\bibinfo
  {journal} {Ann. Phys.}\ }\textbf {\bibinfo {volume} {529}},\ \bibinfo {pages}
  {1600406} (\bibinfo {year} {2017})}\BibitemShut {NoStop}%
\bibitem [{\citenamefont {Fateev}\ \emph {et~al.}(2017)\citenamefont {Fateev},
  \citenamefont {Mashinsky},\ and\ \citenamefont {Popov}}]{Fateev2017}%
  \BibitemOpen
  \bibfield  {author} {\bibinfo {author} {\bibfnamefont {D.~V.}\ \bibnamefont
  {Fateev}}, \bibinfo {author} {\bibfnamefont {K.~V.}\ \bibnamefont
  {Mashinsky}},\ and\ \bibinfo {author} {\bibfnamefont {V.~V.}\ \bibnamefont
  {Popov}},\ }\bibfield  {title} {\bibinfo {title} {Terahertz plasmonic
  rectification in a spatially periodic graphene},\ }\href
  {https://doi.org/10.1063/1.4975829} {\bibfield  {journal} {\bibinfo
  {journal} {Appl. Phys. Lett.}\ }\textbf {\bibinfo {volume} {110}},\ \bibinfo
  {pages} {061106} (\bibinfo {year} {2017})}\BibitemShut {NoStop}%
\bibitem [{\citenamefont {Fateev}\ \emph {et~al.}(2019)\citenamefont {Fateev},
  \citenamefont {Mashinsky}, \citenamefont {Sun},\ and\ \citenamefont
  {Popov}}]{Fateev2019}%
  \BibitemOpen
  \bibfield  {author} {\bibinfo {author} {\bibfnamefont {D.~V.}\ \bibnamefont
  {Fateev}}, \bibinfo {author} {\bibfnamefont {K.~V.}\ \bibnamefont
  {Mashinsky}}, \bibinfo {author} {\bibfnamefont {J.~D.}\ \bibnamefont {Sun}},\
  and\ \bibinfo {author} {\bibfnamefont {V.~V.}\ \bibnamefont {Popov}},\
  }\bibfield  {title} {\bibinfo {title} {Enhanced plasmonic rectification of
  terahertz radiation in spatially periodic graphene structures towards the
  charge neutrality point},\ }\href {https://doi.org/10.1016/j.sse.2019.04.004}
  {\bibfield  {journal} {\bibinfo  {journal} {Solid-State Electron.}\ }\textbf
  {\bibinfo {volume} {157}},\ \bibinfo {pages} {20} (\bibinfo {year}
  {2019})}\BibitemShut {NoStop}%
\bibitem [{\citenamefont {Boubanga-Tombet}\ \emph {et~al.}(2020)\citenamefont
  {Boubanga-Tombet}, \citenamefont {Knap}, \citenamefont {Yadav}, \citenamefont
  {Satou}, \citenamefont {But}, \citenamefont {Popov}, \citenamefont
  {Gorbenko}, \citenamefont {Kachorovskii},\ and\ \citenamefont
  {Otsuji}}]{BoubangaTombet2020}%
  \BibitemOpen
  \bibfield  {author} {\bibinfo {author} {\bibfnamefont {S.}~\bibnamefont
  {Boubanga-Tombet}}, \bibinfo {author} {\bibfnamefont {W.}~\bibnamefont
  {Knap}}, \bibinfo {author} {\bibfnamefont {D.}~\bibnamefont {Yadav}},
  \bibinfo {author} {\bibfnamefont {A.}~\bibnamefont {Satou}}, \bibinfo
  {author} {\bibfnamefont {D.~B.}\ \bibnamefont {But}}, \bibinfo {author}
  {\bibfnamefont {V.~V.}\ \bibnamefont {Popov}}, \bibinfo {author}
  {\bibfnamefont {I.~V.}\ \bibnamefont {Gorbenko}}, \bibinfo {author}
  {\bibfnamefont {V.}~\bibnamefont {Kachorovskii}},\ and\ \bibinfo {author}
  {\bibfnamefont {T.}~\bibnamefont {Otsuji}},\ }\bibfield  {title} {\bibinfo
  {title} {Room-temperature amplification of terahertz radiation by
  grating-gate graphene structures},\ }\href
  {https://doi.org/10.1103/physrevx.10.031004} {\bibfield  {journal} {\bibinfo
  {journal} {Phys. Rev. X}\ }\textbf {\bibinfo {volume} {10}},\ \bibinfo
  {pages} {031004} (\bibinfo {year} {2020})}\BibitemShut {NoStop}%
\bibitem [{\citenamefont {Delgado-Notario}\ \emph {et~al.}(2020)\citenamefont
  {Delgado-Notario}, \citenamefont {Cleric{\`{o}}}, \citenamefont {Diez},
  \citenamefont {Vel{\'{a}}zquez-P{\'{e}}rez}, \citenamefont {Taniguchi},
  \citenamefont {Watanabe}, \citenamefont {Otsuji},\ and\ \citenamefont
  {Meziani}}]{DelgadoNotario2020}%
  \BibitemOpen
  \bibfield  {author} {\bibinfo {author} {\bibfnamefont {J.~A.}\ \bibnamefont
  {Delgado-Notario}}, \bibinfo {author} {\bibfnamefont {V.}~\bibnamefont
  {Cleric{\`{o}}}}, \bibinfo {author} {\bibfnamefont {E.}~\bibnamefont {Diez}},
  \bibinfo {author} {\bibfnamefont {J.~E.}\ \bibnamefont
  {Vel{\'{a}}zquez-P{\'{e}}rez}}, \bibinfo {author} {\bibfnamefont
  {T.}~\bibnamefont {Taniguchi}}, \bibinfo {author} {\bibfnamefont
  {K.}~\bibnamefont {Watanabe}}, \bibinfo {author} {\bibfnamefont
  {T.}~\bibnamefont {Otsuji}},\ and\ \bibinfo {author} {\bibfnamefont {Y.~M.}\
  \bibnamefont {Meziani}},\ }\bibfield  {title} {\bibinfo {title} {Asymmetric
  dual-grating gates graphene {FET} for detection of terahertz radiations},\
  }\href {https://doi.org/10.1063/5.0007249} {\bibfield  {journal} {\bibinfo
  {journal} {APL Photonics}\ }\textbf {\bibinfo {volume} {5}},\ \bibinfo
  {pages} {066102} (\bibinfo {year} {2020})}\BibitemShut {NoStop}%
\bibitem [{\citenamefont {Delgado-Notario}\ \emph {et~al.}(2022)\citenamefont
  {Delgado-Notario}, \citenamefont {Knap}, \citenamefont {Cleric{\`{o}}},
  \citenamefont {Salvador-S{\'{a}}nchez}, \citenamefont {Calvo-Gallego},
  \citenamefont {Taniguchi}, \citenamefont {Watanabe}, \citenamefont {Otsuji},
  \citenamefont {Popov}, \citenamefont {Fateev}, \citenamefont {Diez},
  \citenamefont {Vel{\'{a}}zquez-P{\'{e}}rez},\ and\ \citenamefont
  {Meziani}}]{DelgadoNotario2022}%
  \BibitemOpen
  \bibfield  {author} {\bibinfo {author} {\bibfnamefont {J.~A.}\ \bibnamefont
  {Delgado-Notario}}, \bibinfo {author} {\bibfnamefont {W.}~\bibnamefont
  {Knap}}, \bibinfo {author} {\bibfnamefont {V.}~\bibnamefont {Cleric{\`{o}}}},
  \bibinfo {author} {\bibfnamefont {J.}~\bibnamefont {Salvador-S{\'{a}}nchez}},
  \bibinfo {author} {\bibfnamefont {J.}~\bibnamefont {Calvo-Gallego}}, \bibinfo
  {author} {\bibfnamefont {T.}~\bibnamefont {Taniguchi}}, \bibinfo {author}
  {\bibfnamefont {K.}~\bibnamefont {Watanabe}}, \bibinfo {author}
  {\bibfnamefont {T.}~\bibnamefont {Otsuji}}, \bibinfo {author} {\bibfnamefont
  {V.~V.}\ \bibnamefont {Popov}}, \bibinfo {author} {\bibfnamefont {D.~V.}\
  \bibnamefont {Fateev}}, \bibinfo {author} {\bibfnamefont {E.}~\bibnamefont
  {Diez}}, \bibinfo {author} {\bibfnamefont {J.~E.}\ \bibnamefont
  {Vel{\'{a}}zquez-P{\'{e}}rez}},\ and\ \bibinfo {author} {\bibfnamefont
  {Y.~M.}\ \bibnamefont {Meziani}},\ }\bibfield  {title} {\bibinfo {title}
  {Enhanced terahertz detection of multigate graphene nanostructures},\ }\href
  {https://doi.org/10.1515/nanoph-2021-0573} {\bibfield  {journal} {\bibinfo
  {journal} {Nanophotonics}\ }\textbf {\bibinfo {volume} {11}},\ \bibinfo
  {pages} {519} (\bibinfo {year} {2022})}\BibitemShut {NoStop}%
\bibitem [{\citenamefont {Morozov}\ \emph {et~al.}(2021)\citenamefont
  {Morozov}, \citenamefont {Popov},\ and\ \citenamefont
  {Fateev}}]{Morozov2021}%
  \BibitemOpen
  \bibfield  {author} {\bibinfo {author} {\bibfnamefont {M.~Y.}\ \bibnamefont
  {Morozov}}, \bibinfo {author} {\bibfnamefont {V.~V.}\ \bibnamefont {Popov}},\
  and\ \bibinfo {author} {\bibfnamefont {D.~V.}\ \bibnamefont {Fateev}},\
  }\bibfield  {title} {\bibinfo {title} {Electrically controllable active
  plasmonic directional coupler of terahertz signal based on a periodical dual
  grating gate graphene structure},\ }\bibfield  {journal} {\bibinfo  {journal}
  {Sci. Rep.}\ }\textbf {\bibinfo {volume} {11}},\ \href
  {https://doi.org/10.1038/s41598-021-90876-2} {10.1038/s41598-021-90876-2}
  (\bibinfo {year} {2021})\BibitemShut {NoStop}%
\bibitem [{\citenamefont {M{\"o}nch}\ \emph {et~al.}(2022)\citenamefont
  {M{\"o}nch}, \citenamefont {Potashin}, \citenamefont {Lindner}, \citenamefont
  {Yahniuk}, \citenamefont {Golub}, \citenamefont {Kachorovskii}, \citenamefont
  {Bel'kov}, \citenamefont {Huber}, \citenamefont {Watanabe}, \citenamefont
  {Taniguchi}, \citenamefont {Eroms}, \citenamefont {Weiss},\ and\
  \citenamefont {Ganichev}}]{Moench2022}%
  \BibitemOpen
  \bibfield  {author} {\bibinfo {author} {\bibfnamefont {E.}~\bibnamefont
  {M{\"o}nch}}, \bibinfo {author} {\bibfnamefont {S.~O.}\ \bibnamefont
  {Potashin}}, \bibinfo {author} {\bibfnamefont {K.}~\bibnamefont {Lindner}},
  \bibinfo {author} {\bibfnamefont {I.}~\bibnamefont {Yahniuk}}, \bibinfo
  {author} {\bibfnamefont {L.~E.}\ \bibnamefont {Golub}}, \bibinfo {author}
  {\bibfnamefont {V.~Y.}\ \bibnamefont {Kachorovskii}}, \bibinfo {author}
  {\bibfnamefont {V.~V.}\ \bibnamefont {Bel'kov}}, \bibinfo {author}
  {\bibfnamefont {R.}~\bibnamefont {Huber}}, \bibinfo {author} {\bibfnamefont
  {K.}~\bibnamefont {Watanabe}}, \bibinfo {author} {\bibfnamefont
  {T.}~\bibnamefont {Taniguchi}}, \bibinfo {author} {\bibfnamefont
  {J.}~\bibnamefont {Eroms}}, \bibinfo {author} {\bibfnamefont
  {D.}~\bibnamefont {Weiss}},\ and\ \bibinfo {author} {\bibfnamefont {S.~D.}\
  \bibnamefont {Ganichev}},\ }\bibfield  {title} {\bibinfo {title} {Ratchet
  effect in spatially modulated bilayer graphene: Signature of hydrodynamic
  transport},\ }\href {https://doi.org/10.1103/physrevb.105.045404} {\bibfield
  {journal} {\bibinfo  {journal} {Phys. Rev. B}\ }\textbf {\bibinfo {volume}
  {105}},\ \bibinfo {pages} {045404} (\bibinfo {year} {2022})}\BibitemShut
  {NoStop}%
\bibitem [{\citenamefont {M{\"o}nch}\ \emph {et~al.}(2023)\citenamefont
  {M{\"o}nch}, \citenamefont {Hubmann}, \citenamefont {Yahniuk}, \citenamefont
  {Schweiss}, \citenamefont {Bel’kov}, \citenamefont {Golub}, \citenamefont
  {Huber}, \citenamefont {Eroms}, \citenamefont {Watanabe}, \citenamefont
  {Taniguchi}, \citenamefont {Weiss},\ and\ \citenamefont
  {Ganichev}}]{Moench2023}%
  \BibitemOpen
  \bibfield  {author} {\bibinfo {author} {\bibfnamefont {E.}~\bibnamefont
  {M{\"o}nch}}, \bibinfo {author} {\bibfnamefont {S.}~\bibnamefont {Hubmann}},
  \bibinfo {author} {\bibfnamefont {I.}~\bibnamefont {Yahniuk}}, \bibinfo
  {author} {\bibfnamefont {S.}~\bibnamefont {Schweiss}}, \bibinfo {author}
  {\bibfnamefont {V.~V.}\ \bibnamefont {Bel’kov}}, \bibinfo {author}
  {\bibfnamefont {L.~E.}\ \bibnamefont {Golub}}, \bibinfo {author}
  {\bibfnamefont {R.}~\bibnamefont {Huber}}, \bibinfo {author} {\bibfnamefont
  {J.}~\bibnamefont {Eroms}}, \bibinfo {author} {\bibfnamefont
  {K.}~\bibnamefont {Watanabe}}, \bibinfo {author} {\bibfnamefont
  {T.}~\bibnamefont {Taniguchi}}, \bibinfo {author} {\bibfnamefont
  {D.}~\bibnamefont {Weiss}},\ and\ \bibinfo {author} {\bibfnamefont {S.~D.}\
  \bibnamefont {Ganichev}},\ }\bibfield  {title} {\bibinfo {title} {Nonlinear
  intensity dependence of ratchet currents induced by terahertz laser radiation
  in bilayer graphene with asymmetric periodic grating gates},\ }\href
  {https://doi.org/10.1063/5.0165248} {\bibfield  {journal} {\bibinfo
  {journal} {J. Appl. Phys.}\ }\textbf {\bibinfo {volume} {134}},\ \bibinfo
  {pages} {123102} (\bibinfo {year} {2023})}\BibitemShut {NoStop}%
\bibitem [{\citenamefont {Hild}\ \emph {et~al.}(2024)\citenamefont {Hild},
  \citenamefont {Yahniuk}, \citenamefont {Golub}, \citenamefont {Amann},
  \citenamefont {Eroms}, \citenamefont {Weiss}, \citenamefont {Watanabe},
  \citenamefont {Taniguchi},\ and\ \citenamefont {Ganichev}}]{Hild2024}%
  \BibitemOpen
  \bibfield  {author} {\bibinfo {author} {\bibfnamefont {M.}~\bibnamefont
  {Hild}}, \bibinfo {author} {\bibfnamefont {I.}~\bibnamefont {Yahniuk}},
  \bibinfo {author} {\bibfnamefont {L.~E.}\ \bibnamefont {Golub}}, \bibinfo
  {author} {\bibfnamefont {J.}~\bibnamefont {Amann}}, \bibinfo {author}
  {\bibfnamefont {J.}~\bibnamefont {Eroms}}, \bibinfo {author} {\bibfnamefont
  {D.}~\bibnamefont {Weiss}}, \bibinfo {author} {\bibfnamefont
  {K.}~\bibnamefont {Watanabe}}, \bibinfo {author} {\bibfnamefont
  {T.}~\bibnamefont {Taniguchi}},\ and\ \bibinfo {author} {\bibfnamefont
  {S.~D.}\ \bibnamefont {Ganichev}},\ }\bibfield  {title} {\bibinfo {title}
  {Circular terahertz ratchets in a two-dimensionally modulated dirac system},\
  }\href {https://doi.org/10.1103/physrevresearch.6.023308} {\bibfield
  {journal} {\bibinfo  {journal} {Phys. Rev. Research}\ }\textbf {\bibinfo
  {volume} {6}},\ \bibinfo {pages} {023308} (\bibinfo {year}
  {2024})}\BibitemShut {NoStop}%
\bibitem [{\citenamefont {Yahniuk}\ \emph {et~al.}(2024)\citenamefont
  {Yahniuk}, \citenamefont {Hild}, \citenamefont {Golub}, \citenamefont
  {Amann}, \citenamefont {Eroms}, \citenamefont {Weiss}, \citenamefont {Kang},
  \citenamefont {Liu}, \citenamefont {Watanabe}, \citenamefont {Taniguchi},\
  and\ \citenamefont {Ganichev}}]{Yahniuk2024}%
  \BibitemOpen
  \bibfield  {author} {\bibinfo {author} {\bibfnamefont {I.}~\bibnamefont
  {Yahniuk}}, \bibinfo {author} {\bibfnamefont {M.}~\bibnamefont {Hild}},
  \bibinfo {author} {\bibfnamefont {L.~E.}\ \bibnamefont {Golub}}, \bibinfo
  {author} {\bibfnamefont {J.}~\bibnamefont {Amann}}, \bibinfo {author}
  {\bibfnamefont {J.}~\bibnamefont {Eroms}}, \bibinfo {author} {\bibfnamefont
  {D.}~\bibnamefont {Weiss}}, \bibinfo {author} {\bibfnamefont {W.-H.}\
  \bibnamefont {Kang}}, \bibinfo {author} {\bibfnamefont {M.-H.}\ \bibnamefont
  {Liu}}, \bibinfo {author} {\bibfnamefont {K.}~\bibnamefont {Watanabe}},
  \bibinfo {author} {\bibfnamefont {T.}~\bibnamefont {Taniguchi}},\ and\
  \bibinfo {author} {\bibfnamefont {S.~D.}\ \bibnamefont {Ganichev}},\
  }\bibfield  {title} {\bibinfo {title} {Terahertz ratchet in graphene
  two-dimensional metamaterial formed by a patterned gate with an antidot
  array},\ }\href {https://doi.org/10.1103/physrevb.109.235428} {\bibfield
  {journal} {\bibinfo  {journal} {Phys. Rev. B}\ }\textbf {\bibinfo {volume}
  {109}},\ \bibinfo {pages} {235428} (\bibinfo {year} {2024})}\BibitemShut
  {NoStop}%
\bibitem [{\citenamefont {Ganichev}(1999)}]{Ganichev1999}%
  \BibitemOpen
  \bibfield  {author} {\bibinfo {author} {\bibfnamefont {S.~D.}\ \bibnamefont
  {Ganichev}},\ }\bibfield  {title} {\bibinfo {title} {Tunnel ionization of
  deep impurities in semiconductors induced by terahertz electric fields},\
  }\href {https://doi.org/10.1016/s0921-4526(99)00637-7} {\bibfield  {journal}
  {\bibinfo  {journal} {Phys. B}\ }\textbf {\bibinfo {volume} {273-274}},\
  \bibinfo {pages} {737} (\bibinfo {year} {1999})}\BibitemShut {NoStop}%
\bibitem [{\citenamefont {Herrmann}\ \emph {et~al.}(2016)\citenamefont
  {Herrmann}, \citenamefont {Dmitriev}, \citenamefont {Kozlov}, \citenamefont
  {Schneider}, \citenamefont {Jentzsch}, \citenamefont {Kvon}, \citenamefont
  {Olbrich}, \citenamefont {Bel'kov}, \citenamefont {Bayer}, \citenamefont
  {Schuh}, \citenamefont {Bougeard}, \citenamefont {Kuczmik}, \citenamefont
  {Oltscher}, \citenamefont {Weiss},\ and\ \citenamefont
  {Ganichev}}]{Herrmann2016}%
  \BibitemOpen
  \bibfield  {author} {\bibinfo {author} {\bibfnamefont {T.}~\bibnamefont
  {Herrmann}}, \bibinfo {author} {\bibfnamefont {I.~A.}\ \bibnamefont
  {Dmitriev}}, \bibinfo {author} {\bibfnamefont {D.~A.}\ \bibnamefont
  {Kozlov}}, \bibinfo {author} {\bibfnamefont {M.}~\bibnamefont {Schneider}},
  \bibinfo {author} {\bibfnamefont {B.}~\bibnamefont {Jentzsch}}, \bibinfo
  {author} {\bibfnamefont {Z.~D.}\ \bibnamefont {Kvon}}, \bibinfo {author}
  {\bibfnamefont {P.}~\bibnamefont {Olbrich}}, \bibinfo {author} {\bibfnamefont
  {V.~V.}\ \bibnamefont {Bel'kov}}, \bibinfo {author} {\bibfnamefont
  {A.}~\bibnamefont {Bayer}}, \bibinfo {author} {\bibfnamefont
  {D.}~\bibnamefont {Schuh}}, \bibinfo {author} {\bibfnamefont
  {D.}~\bibnamefont {Bougeard}}, \bibinfo {author} {\bibfnamefont
  {T.}~\bibnamefont {Kuczmik}}, \bibinfo {author} {\bibfnamefont
  {M.}~\bibnamefont {Oltscher}}, \bibinfo {author} {\bibfnamefont
  {D.}~\bibnamefont {Weiss}},\ and\ \bibinfo {author} {\bibfnamefont {S.~D.}\
  \bibnamefont {Ganichev}},\ }\bibfield  {title} {\bibinfo {title} {Analog of
  microwave-induced resistance oscillations induced in {GaAs} heterostructures
  by terahertz radiation},\ }\href {https://doi.org/10.1103/physrevb.94.081301}
  {\bibfield  {journal} {\bibinfo  {journal} {Phys. Rev. B}\ }\textbf {\bibinfo
  {volume} {94}},\ \bibinfo {pages} {081301} (\bibinfo {year}
  {2016})}\BibitemShut {NoStop}%
\bibitem [{\citenamefont {Glazov}\ and\ \citenamefont
  {Ganichev}(2014)}]{Glazov2014}%
  \BibitemOpen
  \bibfield  {author} {\bibinfo {author} {\bibfnamefont {M.~M.}\ \bibnamefont
  {Glazov}}\ and\ \bibinfo {author} {\bibfnamefont {S.~D.}\ \bibnamefont
  {Ganichev}},\ }\bibfield  {title} {\bibinfo {title} {High frequency electric
  field induced nonlinear effects in graphene},\ }\href
  {https://doi.org/10.1016/j.physrep.2013.10.003} {\bibfield  {journal}
  {\bibinfo  {journal} {Phys. Rep.}\ }\textbf {\bibinfo {volume} {535}},\
  \bibinfo {pages} {101} (\bibinfo {year} {2014})}\BibitemShut {NoStop}%
\bibitem [{\citenamefont {Hubmann}\ \emph
  {et~al.}(2020{\natexlab{a}})\citenamefont {Hubmann}, \citenamefont {Budkin},
  \citenamefont {Otteneder}, \citenamefont {But}, \citenamefont {Sacre},
  \citenamefont {Yahniuk}, \citenamefont {Diendorfer}, \citenamefont {Belkov},
  \citenamefont {Kozlov}, \citenamefont {Mikhailov}, \citenamefont {Dvoretsky},
  \citenamefont {Varavin}, \citenamefont {Remesnik}, \citenamefont {Tarasenko},
  \citenamefont {Knap},\ and\ \citenamefont {Ganichev}}]{Hubmann2020}%
  \BibitemOpen
  \bibfield  {author} {\bibinfo {author} {\bibfnamefont {S.}~\bibnamefont
  {Hubmann}}, \bibinfo {author} {\bibfnamefont {G.~V.}\ \bibnamefont {Budkin}},
  \bibinfo {author} {\bibfnamefont {M.}~\bibnamefont {Otteneder}}, \bibinfo
  {author} {\bibfnamefont {D.}~\bibnamefont {But}}, \bibinfo {author}
  {\bibfnamefont {D.}~\bibnamefont {Sacre}}, \bibinfo {author} {\bibfnamefont
  {I.}~\bibnamefont {Yahniuk}}, \bibinfo {author} {\bibfnamefont
  {K.}~\bibnamefont {Diendorfer}}, \bibinfo {author} {\bibfnamefont {V.~V.}\
  \bibnamefont {Belkov}}, \bibinfo {author} {\bibfnamefont {D.~A.}\
  \bibnamefont {Kozlov}}, \bibinfo {author} {\bibfnamefont {N.~N.}\
  \bibnamefont {Mikhailov}}, \bibinfo {author} {\bibfnamefont {S.~A.}\
  \bibnamefont {Dvoretsky}}, \bibinfo {author} {\bibfnamefont {V.~S.}\
  \bibnamefont {Varavin}}, \bibinfo {author} {\bibfnamefont {V.~G.}\
  \bibnamefont {Remesnik}}, \bibinfo {author} {\bibfnamefont {S.~A.}\
  \bibnamefont {Tarasenko}}, \bibinfo {author} {\bibfnamefont {W.}~\bibnamefont
  {Knap}},\ and\ \bibinfo {author} {\bibfnamefont {S.~D.}\ \bibnamefont
  {Ganichev}},\ }\bibfield  {title} {\bibinfo {title} {Symmetry breaking and
  circular photogalvanic effect in epitaxial cdhg films},\ }\href
  {https://doi.org/10.1103/physrevmaterials.4.043607} {\bibfield  {journal}
  {\bibinfo  {journal} {Physical Review Materials}\ }\textbf {\bibinfo {volume}
  {4}},\ \bibinfo {pages} {043607} (\bibinfo {year}
  {2020}{\natexlab{a}})}\BibitemShut {NoStop}%
\bibitem [{\citenamefont {Hubmann}\ \emph
  {et~al.}(2020{\natexlab{b}})\citenamefont {Hubmann}, \citenamefont {Bel'kov},
  \citenamefont {Golub}, \citenamefont {Kachorovskii}, \citenamefont
  {Drienovsky}, \citenamefont {Eroms}, \citenamefont {Weiss},\ and\
  \citenamefont {Ganichev}}]{Hubmann2020a}%
  \BibitemOpen
  \bibfield  {author} {\bibinfo {author} {\bibfnamefont {S.}~\bibnamefont
  {Hubmann}}, \bibinfo {author} {\bibfnamefont {V.~V.}\ \bibnamefont
  {Bel'kov}}, \bibinfo {author} {\bibfnamefont {L.~E.}\ \bibnamefont {Golub}},
  \bibinfo {author} {\bibfnamefont {V.~Y.}\ \bibnamefont {Kachorovskii}},
  \bibinfo {author} {\bibfnamefont {M.}~\bibnamefont {Drienovsky}}, \bibinfo
  {author} {\bibfnamefont {J.}~\bibnamefont {Eroms}}, \bibinfo {author}
  {\bibfnamefont {D.}~\bibnamefont {Weiss}},\ and\ \bibinfo {author}
  {\bibfnamefont {S.~D.}\ \bibnamefont {Ganichev}},\ }\bibfield  {title}
  {\bibinfo {title} {Giant ratchet magneto-photocurrent in graphene lateral
  superlattices},\ }\href {https://doi.org/10.1103/physrevresearch.2.033186}
  {\bibfield  {journal} {\bibinfo  {journal} {Phys. Rev. Research}\ }\textbf
  {\bibinfo {volume} {2}},\ \bibinfo {pages} {033186} (\bibinfo {year}
  {2020}{\natexlab{b}})}\BibitemShut {NoStop}%
\bibitem [{\citenamefont {Budkin}\ \emph {et~al.}(2016)\citenamefont {Budkin},
  \citenamefont {Golub}, \citenamefont {Ivchenko},\ and\ \citenamefont
  {Ganichev}}]{Budkin2016a}%
  \BibitemOpen
  \bibfield  {author} {\bibinfo {author} {\bibfnamefont {G.~V.}\ \bibnamefont
  {Budkin}}, \bibinfo {author} {\bibfnamefont {L.~E.}\ \bibnamefont {Golub}},
  \bibinfo {author} {\bibfnamefont {E.~L.}\ \bibnamefont {Ivchenko}},\ and\
  \bibinfo {author} {\bibfnamefont {S.~D.}\ \bibnamefont {Ganichev}},\
  }\bibfield  {title} {\bibinfo {title} {Magnetic ratchet effects in a
  two-dimensional electron gas},\ }\href
  {https://doi.org/10.1134/s0021364016210074} {\bibfield  {journal} {\bibinfo
  {journal} {JETP Lett.}\ }\textbf {\bibinfo {volume} {104}},\ \bibinfo {pages}
  {649} (\bibinfo {year} {2016})}\BibitemShut {NoStop}%
\bibitem [{\citenamefont {Mönch}\ \emph {et~al.}(2023)\citenamefont {Mönch},
  \citenamefont {Potashin}, \citenamefont {Lindner}, \citenamefont {Yahniuk},
  \citenamefont {Golub}, \citenamefont {Kachorovskii}, \citenamefont
  {Bel’kov}, \citenamefont {Huber}, \citenamefont {Watanabe}, \citenamefont
  {Taniguchi}, \citenamefont {Eroms}, \citenamefont {Weiss},\ and\
  \citenamefont {Ganichev}}]{Moench2023b}%
  \BibitemOpen
  \bibfield  {author} {\bibinfo {author} {\bibfnamefont {E.}~\bibnamefont
  {Mönch}}, \bibinfo {author} {\bibfnamefont {S.~O.}\ \bibnamefont
  {Potashin}}, \bibinfo {author} {\bibfnamefont {K.}~\bibnamefont {Lindner}},
  \bibinfo {author} {\bibfnamefont {I.}~\bibnamefont {Yahniuk}}, \bibinfo
  {author} {\bibfnamefont {L.~E.}\ \bibnamefont {Golub}}, \bibinfo {author}
  {\bibfnamefont {V.~Y.}\ \bibnamefont {Kachorovskii}}, \bibinfo {author}
  {\bibfnamefont {V.~V.}\ \bibnamefont {Bel’kov}}, \bibinfo {author}
  {\bibfnamefont {R.}~\bibnamefont {Huber}}, \bibinfo {author} {\bibfnamefont
  {K.}~\bibnamefont {Watanabe}}, \bibinfo {author} {\bibfnamefont
  {T.}~\bibnamefont {Taniguchi}}, \bibinfo {author} {\bibfnamefont
  {J.}~\bibnamefont {Eroms}}, \bibinfo {author} {\bibfnamefont
  {D.}~\bibnamefont {Weiss}},\ and\ \bibinfo {author} {\bibfnamefont {S.~D.}\
  \bibnamefont {Ganichev}},\ }\bibfield  {title} {\bibinfo {title} {Cyclotron
  and magnetoplasmon resonances in bilayer graphene ratchets},\ }\href
  {https://doi.org/10.1103/physrevb.107.115408} {\bibfield  {journal} {\bibinfo
   {journal} {Phys. Rev. B}\ }\textbf {\bibinfo {volume} {107}},\ \bibinfo
  {pages} {115408} (\bibinfo {year} {2023})}\BibitemShut {NoStop}%
\bibitem [{\citenamefont {Bray}\ \emph {et~al.}(2022)\citenamefont {Bray},
  \citenamefont {Maussang}, \citenamefont {Consejo}, \citenamefont
  {Delgado-Notario}, \citenamefont {Krishtopenko}, \citenamefont {Yahniuk},
  \citenamefont {Gebert}, \citenamefont {Ruffenach}, \citenamefont {Dinar},
  \citenamefont {Moench}, \citenamefont {Eroms}, \citenamefont {Indykiewicz},
  \citenamefont {Jouault}, \citenamefont {Torres}, \citenamefont {Meziani},
  \citenamefont {Knap}, \citenamefont {Yurgens}, \citenamefont {Ganichev},\
  and\ \citenamefont {Teppe}}]{Bray2022}%
  \BibitemOpen
  \bibfield  {author} {\bibinfo {author} {\bibfnamefont {C.}~\bibnamefont
  {Bray}}, \bibinfo {author} {\bibfnamefont {K.}~\bibnamefont {Maussang}},
  \bibinfo {author} {\bibfnamefont {C.}~\bibnamefont {Consejo}}, \bibinfo
  {author} {\bibfnamefont {J.~A.}\ \bibnamefont {Delgado-Notario}}, \bibinfo
  {author} {\bibfnamefont {S.}~\bibnamefont {Krishtopenko}}, \bibinfo {author}
  {\bibfnamefont {I.}~\bibnamefont {Yahniuk}}, \bibinfo {author} {\bibfnamefont
  {S.}~\bibnamefont {Gebert}}, \bibinfo {author} {\bibfnamefont
  {S.}~\bibnamefont {Ruffenach}}, \bibinfo {author} {\bibfnamefont
  {K.}~\bibnamefont {Dinar}}, \bibinfo {author} {\bibfnamefont
  {E.}~\bibnamefont {Moench}}, \bibinfo {author} {\bibfnamefont
  {J.}~\bibnamefont {Eroms}}, \bibinfo {author} {\bibfnamefont
  {K.}~\bibnamefont {Indykiewicz}}, \bibinfo {author} {\bibfnamefont
  {B.}~\bibnamefont {Jouault}}, \bibinfo {author} {\bibfnamefont
  {J.}~\bibnamefont {Torres}}, \bibinfo {author} {\bibfnamefont {Y.~M.}\
  \bibnamefont {Meziani}}, \bibinfo {author} {\bibfnamefont {W.}~\bibnamefont
  {Knap}}, \bibinfo {author} {\bibfnamefont {A.}~\bibnamefont {Yurgens}},
  \bibinfo {author} {\bibfnamefont {S.~D.}\ \bibnamefont {Ganichev}},\ and\
  \bibinfo {author} {\bibfnamefont {F.}~\bibnamefont {Teppe}},\ }\bibfield
  {title} {\bibinfo {title} {Temperature-dependent zero-field splittings in
  graphene},\ }\href {https://doi.org/10.1103/physrevb.106.245141} {\bibfield
  {journal} {\bibinfo  {journal} {Physical Review B}\ }\textbf {\bibinfo
  {volume} {106}},\ \bibinfo {pages} {245141} (\bibinfo {year}
  {2022})}\BibitemShut {NoStop}%
\bibitem [{\citenamefont {Olbrich}\ \emph {et~al.}(2009)\citenamefont
  {Olbrich}, \citenamefont {Ivchenko}, \citenamefont {Ravash}, \citenamefont
  {Feil}, \citenamefont {Danilov}, \citenamefont {Allerdings}, \citenamefont
  {Weiss}, \citenamefont {Schuh}, \citenamefont {Wegscheider},\ and\
  \citenamefont {Ganichev}}]{Olbrich2009ratchet}%
  \BibitemOpen
  \bibfield  {author} {\bibinfo {author} {\bibfnamefont {P.}~\bibnamefont
  {Olbrich}}, \bibinfo {author} {\bibfnamefont {E.~L.}\ \bibnamefont
  {Ivchenko}}, \bibinfo {author} {\bibfnamefont {R.}~\bibnamefont {Ravash}},
  \bibinfo {author} {\bibfnamefont {T.}~\bibnamefont {Feil}}, \bibinfo {author}
  {\bibfnamefont {S.~D.}\ \bibnamefont {Danilov}}, \bibinfo {author}
  {\bibfnamefont {J.}~\bibnamefont {Allerdings}}, \bibinfo {author}
  {\bibfnamefont {D.}~\bibnamefont {Weiss}}, \bibinfo {author} {\bibfnamefont
  {D.}~\bibnamefont {Schuh}}, \bibinfo {author} {\bibfnamefont
  {W.}~\bibnamefont {Wegscheider}},\ and\ \bibinfo {author} {\bibfnamefont
  {S.~D.}\ \bibnamefont {Ganichev}},\ }\bibfield  {title} {\bibinfo {title}
  {Ratchet effects induced by terahertz radiation in heterostructures with a
  lateral periodic potential},\ }\href
  {https://doi.org/10.1103/physrevlett.103.090603} {\bibfield  {journal}
  {\bibinfo  {journal} {Phys. Rev. Lett.}\ }\textbf {\bibinfo {volume} {103}},\
  \bibinfo {pages} {090603} (\bibinfo {year} {2009})}\BibitemShut {NoStop}%
\bibitem [{\citenamefont {Ivchenko}\ and\ \citenamefont
  {Ganichev}(2011)}]{Ivchenko2011}%
  \BibitemOpen
  \bibfield  {author} {\bibinfo {author} {\bibfnamefont {E.~L.}\ \bibnamefont
  {Ivchenko}}\ and\ \bibinfo {author} {\bibfnamefont {S.~D.}\ \bibnamefont
  {Ganichev}},\ }\bibfield  {title} {\bibinfo {title} {Ratchet effects in
  quantum wells with a lateral superlattice},\ }\href
  {https://doi.org/10.1134/s002136401111004x} {\bibfield  {journal} {\bibinfo
  {journal} {JETP Lett.}\ }\textbf {\bibinfo {volume} {93}},\ \bibinfo {pages}
  {673} (\bibinfo {year} {2011})},\ \bibinfo {note} {[Pisma v ZhETF
  \textbf{93}, 752 (2011)]}\BibitemShut {NoStop}%
\bibitem [{\citenamefont {Nalitov}\ \emph {et~al.}(2012)\citenamefont
  {Nalitov}, \citenamefont {Golub},\ and\ \citenamefont
  {Ivchenko}}]{Nalitov2012}%
  \BibitemOpen
  \bibfield  {author} {\bibinfo {author} {\bibfnamefont {A.~V.}\ \bibnamefont
  {Nalitov}}, \bibinfo {author} {\bibfnamefont {L.~E.}\ \bibnamefont {Golub}},\
  and\ \bibinfo {author} {\bibfnamefont {E.~L.}\ \bibnamefont {Ivchenko}},\
  }\bibfield  {title} {\bibinfo {title} {Ratchet effects in two-dimensional
  systems with a lateral periodic potential},\ }\href
  {https://doi.org/10.1103/physrevb.86.115301} {\bibfield  {journal} {\bibinfo
  {journal} {Phys. Rev. B}\ }\textbf {\bibinfo {volume} {86}},\ \bibinfo
  {pages} {115301} (\bibinfo {year} {2012})}\BibitemShut {NoStop}%
\end{thebibliography}%
 	\end{document}